\documentclass[fleqn,usenatbib]{mnras}

\usepackage{graphicx}
\usepackage{amsmath}
\usepackage[dvipsnames]{xcolor}
\usepackage{amssymb}
\usepackage{newtxtext,newtxmath}
\usepackage{ulem}
\usepackage{siunitx}
\usepackage{macros_vdb}
\usepackage{tikz}
\usetikzlibrary{shapes,arrows}

\usepackage{etoolbox}
\makeatletter
\patchcmd\@combinedblfloats{\box\@outputbox}{\unvbox\@outputbox}{}{\errmessage{\noexpand patch failed}}
\makeatother

\graphicspath{{./}{figures/}}

\defcitealias{Komatsu.Seljak2001}{KS01}
\defcitealias{Shi2014}{SK14}

\title[Impact of MAHs on the $Y_\mathrm{SZ}-M$ scaling relation]
      {Scatter in Sunyaev--Zel'dovich effect scaling relations explained by inter-cluster variance in mass accretion histories}

\author[S. B. Green et al.]{%
        Sheridan~B.~Green,$^{1}$\thanks{E-mail: \href{sheridan.green@yale.edu}{sheridan.green@yale.edu} (SBG)}\thanks{NSF Graduate Research Fellow} Han~Aung,$^{1}$ Daisuke~Nagai,$^{1,2}$ and Frank~C.~van~den~Bosch$^{1,2}$
\vspace*{8pt}
\\
$^{1}$Department of Physics, Yale University, P.O. Box 208120, New Haven, CT 06520, U.S.A.\\
$^{2}$Department of Astronomy, Yale University, P.O. Box 208101, New Haven, CT 06520, U.S.A.\\
}

\date{}

\pubyear{2020}

\begin{document}
\label{firstpage}
\pagerange{\pageref{firstpage}--\pageref{lastpage}}

\maketitle

%%%%%%%%%%%%%%%%%%%%%%%%%%%%%%%%%%%%%%%%%%%%%%%%%%%%%%%%%%%%%%%%%%%%%%%%%%

\begin{abstract}
X-ray and microwave cluster scaling relations are immensely valuable for cosmological analysis. However, their power is limited by astrophysical systematics that bias mass estimates and introduce additional scatter. Turbulence injected into the intracluster medium via mass assembly contributes substantially to cluster non-thermal pressure support, a significant source of such uncertainties. We use an analytical model to compute the assembly-driven non-thermal pressure profiles of haloes based on Monte Carlo-generated accretion histories. We introduce a fitting function for the average non-thermal pressure fraction profile, which exhibits minimal dependence on redshift at fixed peak height. Using the model, we predict deviations from self-similarity and the intrinsic scatter in the Sunyaev--Zel'dovich effect observable-mass scaling relation ($Y_\mathrm{SZ}-M$) due solely to inter-cluster variation in mass accretion histories. We study the dependence of $Y_\mathrm{SZ}-M$ on aperture radius, cosmology, redshift, and mass limit. The model predicts $5-9\%$ scatter in $Y_\mathrm{SZ}-M$ at $z=0$, increasing as the aperture used to compute $Y_\mathrm{SZ}$ increases from $R_\mathrm{500c}$ to $5R_\mathrm{500c}$. The predicted scatter lies slightly below that of studies based on non-radiative hydro-simulations, illustrating that assembly history variance is likely responsible for a substantial fraction of scatter in $Y_\mathrm{SZ}-M$. This should be regarded as a lower bound, which will likely increase with the use of an updated gas density model that incorporates a more realistic response to halo assembly. As redshift increases, $Y_\mathrm{SZ}-M$ deviates more from self-similarity and scatter increases. We show that the $Y_\mathrm{SZ}-M$ residuals correlate strongly with the recent halo mass accretion rate, potentially providing an opportunity to infer the latter.
\end{abstract}

%%%%%%%%%%%%%%%%%%%%%%%%%%%%%%%%%%%%%%%%%%%%%%%%%%%%%%%%%%%%%%%%%%%%%%%%%%


%%%%%%%%%%%%%%%%%%%%%%%%%%%%%%%%%%%%%%%%%%%%%%%%%%%%%%%%%%%%%%%%%%%%%%%%%%

\section{Introduction}\label{sec:intro}

Galaxy clusters are the largest gravitationally bound objects in the Universe, forming hierarchically through accretion at the intersection of cosmic filaments. Their mass- and redshift-distribution is intimately connected to the underlying cosmological model. Hence, a precise approach to linking cluster observables (such as X-ray luminosity or the \citet[SZ;][]{SZ1972} effect in the microwave) to the underlying halo mass is essential for using cluster counts as a cosmological probe \citep[][for a recent review]{Allen2011, Pratt2019}.

In the upcoming years, the observed X-ray and SZ cluster samples are forecast to grow tremendously. In the X-ray, the recently-launched \textit{eROSITA} mission is set to discover ${\gtrsim}10^6$ groups and clusters \citep{Pillepich2018}. In the microwave, the Simons Observatory --- planned to begin observations in the early 2020s --- will detect the SZ signal of ${\gtrsim}10^5$ clusters out to high redshifts \citep{Simons}, a catalog that will eventually be augmented to ${\gtrsim}10^6$ objects by the next-generation CMB-S4 project \citep{CMBS4}. The statistical precision of these surveys will enable unprecedentedly tight cosmological constraints, further stress-testing the standard model of cosmology and potentially illuminating the signal of massive neutrinos or dynamical dark energy. Unlocking the full statistical potential of these surveys necessitates the mitigation of systematic uncertainties associated with cluster gas physics, motivating the development of \new{new halo mass proxies with reduced intrinsic scatter and bias relative to current techniques.}

The X-ray luminosity, $L_X$, is a direct, low-cost mass estimator, but it suffers from high intrinsic scatter due to poorly-understood cluster core physics. This scatter can be reduced via core-excision \citep{Maughan2007, Mantz2018} or modeling \citep{Kafer2020} at the cost of sacrificing a significant fraction of the total X-ray photon distribution that comes from the cluster core regions. The integrated SZ signal, $Y_\mathrm{SZ}$, is predicted to have a low intrinsic scatter \citep[$10-15\%$ at fixed mass, e.g.,][]{Nagai2006, Battaglia2012} and is much less sensitive to cluster core physics, as the SZ signal arises from the gas permeating throughout the virialized region of galaxy clusters. The product of the X-ray core-excised spectral temperature and gas mass, $Y_X$ \citep{Kravtsov2006}, has comparable scatter to $Y_\mathrm{SZ}$, but is only obtainable with high-resolution, long-exposure observations of massive, nearby clusters. 

Both X-ray and SZ mass proxies are also subject to scatter due to inter-cluster variance in halo mass accretion histories \new{\citep[MAHs; e.g.,][]{Hoekstra2012, Krause2012, Barnes2017b}}, which results in the presence of varying levels of non-thermal pressure support \citep{Lau2009, Nelson2014}. However, the X-ray signal is further afflicted by \new{cooling and heating mechanisms \citep{Stanek2010}}, gas clumping \citep[e.g.,][]{Nagai2011, Zhuravleva2013, Khedekar2013}, temperature inhomogeneities \citep[e.g.,][]{Rasia2014}, and the cluster dynamical state \citep[e.g.,][]{Ventimiglia2008}, whereas the SZ signal is expected to be less sensitive to these details \citep[e.g.,][but see also \citealt{Marrone2012}]{Motl2005, Wik2008, Eckert2015}. Recent machine learning-based efforts have illustrated that the scatter can be reduced by accounting for the dynamical state via full X-ray images \citep{Ntampaka2019} or summary statistics of the cluster morphology \citep{Green2019}. \new{Understanding the covariance among these multiple observables will be important for constraining cosmological parameters using multi-wavelength cluster surveys \citep{Stanek2010}}. 

In addition to introducing scatter, non-thermal pressure support is responsible for a substantial bias that adversely impacts X-ray- and SZ-based mass proxies. These masses are typically estimated under the assumption of hydrostatic equilibrium (HSE) between the gravitational potential and the observed thermal pressure, which is used in lieu of the total pressure. Because of this assumption, the presence of non-thermal pressure in the cluster introduces a HSE mass bias, resulting in observed X-ray/SZ-based masses that are up to 30\% lower than the corresponding gravitational lensing-based masses \citep[e.g.,][]{Zhang2010, Mahdavi2013, vdLinden2014, Applegate2014, Hoekstra2015, Medezinski2018, Miyatake2019}. Recent observational studies, however, have shown that this bias is much lower for relaxed populations of clusters that have not recently experienced a significant merger event \citep[e.g.,][]{Applegate2016, Eckert2019, Ettori2019, Ghirardini2019}. 

\new{To date, the HSE mass biases of simulated clusters have been estimated to be $5-40\%$} \citep[e.g.,][]{Evrard1990,Rasia2006, Nagai2007, Lau2009, Battaglia2012, Lau2013, Nelson2014b, Shi2016, Biffi2016, Henson2017, Ansarifard2020, Barnes2020}, revealing that bulk and turbulent intracluster gas motions driven by halo mergers and accretion are likely the dominant source of non-thermal pressure \citep[e.g.,][]{Nelson2012, Avestruz2016, Shi2018, Shi2020} (the potential implications of other sources are discussed in Section~\ref{sec:discussion}). Measurements of optical weak lensing masses via background galaxies \citep[e.g.,][]{Dietrich2019} and CMB lensing masses \citep[e.g.,][]{Raghunathan2019} may provide a method of calibrating the cluster mass scales and mitigating the HSE bias problem.

As we approach the low-noise, high-resolution frontier of CMB survey science \citep{Mroczkowski2019}, the SZ effect offers promising potential as a cosmological probe. In contrast to X-ray mass proxies, the SZ-based approach suffers from fewer astrophysical systematics and has greater sensitivity to high-redshifts and cluster outskirts. However, as discussed above, assembly-driven non-thermal pressure support is a dominant systematic impeding SZ science. Hydrodynamical simulations demonstrate that the cluster outskirts, which contribute the majority of the thermal SZ (tSZ) signal, have non-thermal pressure support similar in magnitude to the thermal pressure \citep[e.g.,][for a recent review]{Nelson2014, Vazza2018, Walker2019}. In addition to contributing to the scatter and bias in $Y_\mathrm{SZ}$-based mass estimation, non-thermal pressure also impacts the tSZ angular power spectrum, $C_l$, which is extremely sensitive to the matter density fluctuation amplitude, $C_l \propto \sigma_8^{7-8}$ \citep{Komatsu2002}. Simulation studies have demonstrated that properly accounting for non-thermal pressure can change the SZ power spectrum amplitude by ${\sim}60\%$ \citep[][]{Battaglia2010, Shaw2010, Trac2011}, impacting constraints on $\sigma_8$ and dark energy \citep{Bolliet2018}. Cross-correlation analyses of SZ, lensing, and galaxy surveys have also been used to constrain the HSE mass bias \citep[e.g.,][]{Makiya2018, Makiya2020, Osato2020} as well as the roles of AGN feedback and non-thermal pressure of the warm-hot diffuse baryons in groups and clusters \citep[e.g.,][]{vanWaerbeke2014, Battaglia2015, Hojjati2017, Osato2018}. Hence, accurately characterizing the average non-thermal pressure profile as a function of cluster mass and redshift is crucial for both subjugating the HSE mass bias problem and using auto- and cross-correlation statistics from upcoming SZ surveys for cosmology. In addition, studying how diversity in halo assembly drives the scatter in the non-thermal pressure support and SZ signal may inform techniques for constructing a more powerful, lower-scatter SZ-based mass proxy that could ultimately strengthen next-generation cosmological analyses.

In this paper, we study analytically the impact of structure formation-generated turbulence on the scatter in the SZ effect observable-mass scaling relation ($Y_\mathrm{SZ}-M$). This is made possible by combining the \citet{Komatsu.Seljak2001} model of the cluster total pressure and gas density profiles, the \citet{Shi2014} analytical model of the mass assembly-driven non-thermal pressure profiles, and both average halo MAHs \citep{vdB2014} and individual Monte Carlo-generated MAHs \citep{Parkinson.etal.08}. Along the way, we identify a near-universality of the average non-thermal pressure fraction profiles, $f_\mathrm{nth}(r)$, at fixed peak height that was first hinted at in the simulations of \citet{Nelson2014}. We then calculate the impact of mass assembly on the HSE mass bias, finding that the average bias should increase considerably with both halo mass and redshift due to larger rates of recent mass accretion. Using the thermal pressure profiles computed for various cluster samples, we investigate the slope, normalization, and intrinsic scatter of the $Y_\mathrm{SZ}-M$ relation and its dependence on aperture radius, redshift, cosmology, and halo mass limit. Importantly, we show that a substantial fraction of the scatter seen in simulated and observed $Y_\mathrm{SZ}-M$ relations can be attributed to inter-cluster variance in the MAHs. Lastly, we identify a strong correlation between the $Y_\mathrm{SZ}-M$ residuals and the recent halo mass accretion rate over the previous dynamical time, a relationship that may enable estimation of the accretion rate in observed clusters. 

This paper is organized as follows. In Section \ref{sec:methods}, we describe our methodology, briefly reviewing models of the cluster gas and thermal pressure profiles and the MAHs as well as defining our observables of interest. In Section \ref{sec:results}, we lay out the results of our analyses, including the predictions for cluster non-thermal pressure profiles (Section \ref{ssec:fnth}), HSE mass biases (Section \ref{ssec:hse}), observable-mass relationships (Section \ref{ssec:omr}), and the connection between $Y_\mathrm{SZ}-M$ residuals and the recent halo mass accretion rate (Section \ref{ssec:mar}). We discuss the implications of the model in Section \ref{sec:discussion}, concluding with a summary of our findings and a forecast of future work in Section \ref{sec:conclusion}.

The fiducial cosmology used throughout this work is consistent with the \citet{Planck18} results: $\Omega_\rmm = 0.311$, $\Omega_\Lambda = 0.689$, $\Omega_\rmb h^2 = 0.0224$, $h = 0.677$, $\sigma_8 = 0.810$, and $n_\rms = 0.967$. The base-10 logarithm is denoted by $\log$ and the natural logarithm is denoted by $\ln$. Much of the analysis utilizes the \textsc{colossus} Python package \citep{Diemer2018}.

\section{Methods}\label{sec:methods}

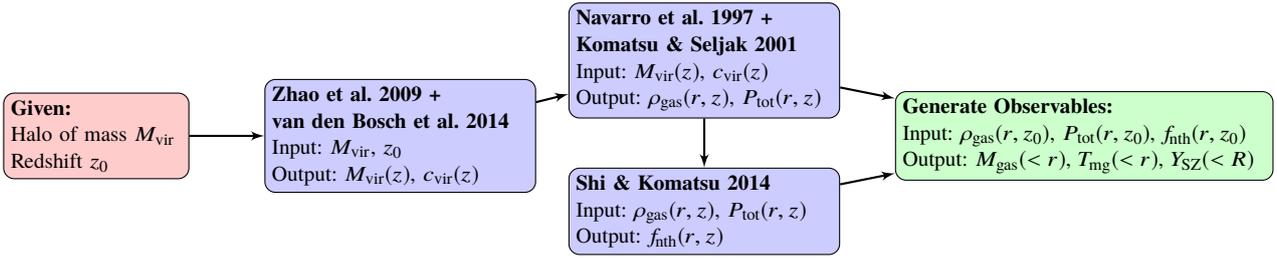
\begin{figure*}
    \centering
    
    \tikzstyle{block} = [rectangle, draw, fill=blue!20,
    text width=12em, rounded corners, minimum height=4em]
    \tikzstyle{blockgreen} = [rectangle, draw, fill=green!20,
    text width=12em, rounded corners, minimum height=4em]
    \tikzstyle{blockred} = [rectangle, draw, fill=red!20,
    text width=12em, rounded corners, minimum height=4em]
    \tikzstyle{line} = [draw, thick, -latex']
    
    \begin{tikzpicture} [node distance = 4cm, auto]
        \node [blockred, text width=8em] (given) {
        \textbf{Given:}\\
        Halo of mass $M_\mathrm{vir}$\\
        Redshift $z_0$};
        \node [block, right of=given] (zhao) {
        \textbf{Zhao et al. 2009 + \\
        van den Bosch et al. 2014}\\
        Input: $M_\mathrm{vir}$, $z_0$ \\
        Output: $M_\mathrm{vir}(z)$, $c_\mathrm{vir}(z)$};
        \node [block, right of=zhao, node distance=4cm, yshift=1cm] (nfw) {
        \textbf{Navarro et al. 1997 + \\
        Komatsu \& Seljak 2001}\\
        Input: $M_\mathrm{vir}(z)$, $c_\mathrm{vir}(z)$ \\
        Output: $\rho_\mathrm{gas}(r,z)$, $P_\mathrm{tot}(r,z)$};
        \node [block, below of=nfw, node distance=2cm] (sk) {
        \textbf{Shi \& Komatsu 2014}\\
        Input: $\rho_\mathrm{gas}(r,z)$, $P_\mathrm{tot}(r,z)$ \\
        Output: $f_\mathrm{nth}(r,z)$};
        \node [blockgreen, text width=17em, right of=nfw, node distance=5cm, yshift=-1cm] (obs) {
        \textbf{Generate Observables:}\\
        Input: $\rho_\mathrm{gas}(r,z_0)$, $P_\mathrm{tot}(r,z_0)$, $f_\mathrm{nth}(r,z_0)$ \\
        Output: $M_\mathrm{gas}(<r)$, $T_\mathrm{mg}(<r)$, $Y_\mathrm{SZ}(<R)$};
        \path [line] (given) -- (zhao);
        \path [line] (zhao) -- (nfw);
        \path [line] (nfw) -- (sk);
        \path [line] (nfw) -- (obs);
        \path [line] (sk) -- (obs);
    \end{tikzpicture}
    \caption{\new{A flowchart that summarizes our theoretical framework. For each halo with a virial mass of $M_\mathrm{vir}$ observed at a redshift of $z_0$, the mass accretion history and concentration history are generated following Section~\ref{ssec:mah}. This is input into the gas model (Section~\ref{ssec:presdens}), which assumes hydrostatic equilibrium, in order to generate the gas density and total pressure profiles throughout the accretion history. The non-thermal pressure fraction profile is then generated following Section~\ref{ssec:nth}. Lastly, the gas density profile and total/non-thermal pressure profiles are used to generate the observables: gas mass, temperature, and integrated SZ signal.}}
    \label{fig:schematic}
\end{figure*}

\new{In this section, we present our analytical framework that we use to model the impact of the assembly history on cluster observables.} We first present the theoretical model of the observable-mass scaling relations (Section~\ref{ssec:selfsim}), which is based on the \citet{Kaiser1986} self-similar model. The cluster observables considered in this study are all functions of the gas density, temperature, and thermal pressure in the intracluster medium (ICM). \new{In Section~\ref{ssec:mah}, we describe the techniques used to generate the MAHs of individual clusters \citep{Cole2000, Parkinson.etal.08} and their population averages \citep{vdB2014}, enabling us to study both mean trends and quantify inter-cluster variance.} We assume that the gas density and \textit{total} pressure are well-described by the model of \citet{Komatsu.Seljak2001}, which we present in Section~\ref{ssec:presdens}. The thermal pressure is obtained from the total by subtracting off the non-thermal component, which we compute using the model of \citet{Shi2014}, presented in Section~\ref{ssec:nth}. We assume throughout that the non-thermal pressure is entirely due to turbulence generated during the cluster's mass assembly. Lastly, we lay out our methods used to compute and quantify the properties of cluster scaling relations in Section~\ref{ssec:quant}. \new{The model framework is summarized in Fig.~\ref{fig:schematic}.}

\subsection{Observables and self-similar scaling relations}\label{ssec:selfsim}

Our main goal is to model the scaling relation between the observable, cylindrically-integrated SZ signal, $Y_\mathrm{SZ}$, and the observationally inferred cluster mass. As discussed below, the SZ signal is proportional to both the cluster gas mass, $M_\mathrm{gas}$, and the gas mass-weighted temperature, $T_\mathrm{mg}$. We therefore also analyze the scaling relations between these quantities and cluster mass. We study the dependence of all of these scaling relations on the aperture radius, $r_\mathrm{ap}$, for which we use multiples of $r_\mathrm{500c}$ and $r_\mathrm{200m}$.\footnote{Note that $r_\mathrm{500c}$ is the radius inside of which the mean density is equal to 500 times the \textit{critical} density, $\rho_\rmc (z)$, whereas within $r_\mathrm{200m}$, the mean density is 200 times the \textit{mean matter} density, $\rho_\rmm (z)$. For cluster mass scales, the virial radius is $r_\mathrm{vir} \approx 2 r_\mathrm{500c} \approx 0.8 r_\mathrm{200m}$ at $z=0$.} In what follows, we use $R$ to denote two-dimensional projected distances and $r$ to denote three-dimensional distances; in particular, $R_\mathrm{ap}$ and $r_{\rm ap}$ are used to indicate the aperture radii used for cylindrically- and spherically-integrated quantities, respectively. The total enclosed halo mass is denoted $M(<r_\mathrm{ap})$. We emphasize that throughout this study we always use the same aperture to compute both the total enclosed halo mass and the cluster observables; however, when studying a cylindrically-integrated observable, we shall still compare it to the spherically enclosed halo mass using the same numerical values for both $r_{\rm ap}$ and $R_{\rm ap}$. The three main observables considered in this paper are $Y_\mathrm{SZ}(<R_{\rm ap})$, $M_\mathrm{gas}(<r_{\rm ap})$, and $T_\mathrm{mg}(<r_{\rm ap})$.\footnote{\new{Note that of $M_\mathrm{gas}$, $T_\mathrm{mg}$, and $Y_\mathrm{SZ}$, only $Y_\mathrm{SZ}$ is computed as a two-dimensional projected quantity in this work --- hence, our $T_\mathrm{mg}$ and $M_\mathrm{gas}$ are not direct observables, but studying these spherically-integrated quantities is still illuminating with regards to understanding the $Y_\mathrm{SZ}-M$ relation.}} Computing these quantities requires a model of the \textit{thermal} pressure and gas density profiles, $P_\mathrm{th}(r)$ and $\rho_\mathrm{gas}(r)$, which we describe in Sections~\ref{ssec:presdens} and~\ref{ssec:nth}, respectively.

The self-similar model developed by \citet{Kaiser1986} is a simple model of cluster evolution based on three assumptions: (i) clusters form from peaks in the initial density field of an Einstein--de Sitter universe with $\Omega_\rmm = 1$, (ii) the amplitude of the primordial density fluctuations varies with spatial scale as a power law, and (iii) processes that impact cluster formation do not introduce additional physical scales to the problem \citep{Kravtsov.Borgani2012}. Under these assumptions, the gravitational collapse of galaxy clusters is \textit{self-similar}. However, various astrophysical processes, such as turbulence, introduce additional physical scales, which result in cluster evolution that deviates from self-similarity. Before we investigate how non-thermal pressure support causes deviations from self-similarity, we briefly describe how the relevant observables (i.e., $M_\mathrm{gas}$, $T_\mathrm{mg}$, and $Y_\mathrm{SZ}$) scale in the \citet{Kaiser1986} model.

Given a particular overdensity definition, cluster mass and radius are interchangeable via $M_{\Delta} = (4\pi/3) \Delta \rho_x(z) r_{\Delta}^3$, where $\Delta$ is the overdensity factor. When haloes are defined with respect to a multiple of the critical density, $\rho_x (z) \equiv \rho_\rmc (z) \propto E^2(z) = \Omega_\rmm (1+z)^3 + \Omega_\Lambda$, whereas when they are defined with respect to the mean matter density, $\rho_x (z) \equiv \rho_\rmm (z) \propto (1+z)^3$. 

The \citet{Kaiser1986} model assumes that the density profile of the gas, $\rho_\mathrm{gas}(r)$, is self-similar and that its normalization is such that, for fixed $r_\mathrm{ap}$, the ratio between the enclosed gas mass, given by
\begin{equation}
    M_\mathrm{gas}(<r_\mathrm{ap}) = 4 \pi \int_0^{r_\mathrm{ap}} \rho_\mathrm{gas}(r) r^2 \rmd r ,
\end{equation}
and the enclosed total mass, $M(<r_{\rm ap})$, is independent of halo mass. As we will see, the halo concentration-mass relation introduces an additional \new{mass-dependence} that causes the gas profile shapes to deviate from self-similarity. However, for large $r_\mathrm{ap}$, the assumption of a fixed $M_\mathrm{gas}(<r_\mathrm{ap}) / M(<r_\mathrm{ap})$ is still reasonable because recent findings in both simulations and observations have found that the cumulative gas mass fraction approaches the cosmic baryon fraction at or below ${\sim}(1-2)r_\mathrm{200m}$ \citep[e.g.,][]{Kravtsov2005, Ettori2006, Planelles2012, Eckert2013, Mantz2014, Morandi2015} for clusters with $M_\mathrm{500c} \gtrsim 10^{14} \, h^{-1} M_\odot$ at $0 \lesssim z \lesssim 1$. This assumption is less realistic for lower mass haloes, where gas depletion due to feedback becomes significant.

The \citet{Kaiser1986} model assumes that the gas is in HSE with the gravitational potential and that the logarithmic slopes of the gas density and thermal pressure profiles are independent of halo mass. Hence, from the HSE equation,
\begin{equation}\label{eqn:hse}
    \frac{1}{\rho_\mathrm{gas}(r)} \frac{\rmd P(r)}{\rmd r} = - \frac{\rmd \Phi(r)}{\rmd r} = \frac{GM(<r)}{r^2}\,,
\end{equation}
where $P(r)=P_\mathrm{th}(r)$ in the absence of non-thermal pressure, and assuming an ideal gas, we have that
\begin{equation}
    M(<r) = -\frac{k_\mathrm{B} \, T(r) \, r}{\mu m_\mathrm{p}\,G} \, \left[\frac{\rmd \ln T}{\rmd \ln r} + \frac{\rmd \ln \rho_\mathrm{gas}}{\rmd \ln r}\right] \,,
\end{equation}
where $\mu m_\mathrm{p}$ is the mean particle mass. This equation can be used to solve for $T(r)$ given $M(<r)$ and $\rho_\mathrm{gas}(r)$, from which we compute the spherically-integrated gas mass-weighted temperature
\begin{equation}
    T_\mathrm{mg}(<r_\mathrm{ap}) = \frac{4 \pi}{M_\mathrm{gas}(< r_\mathrm{ap})} \int_0^{r_\mathrm{ap}} \rho_\mathrm{gas}(r) \, T(r) \, r^2 \, \rmd r\,.
\end{equation}
For the self-similar gas density profile assumed in the \citet{Kaiser1986} model, this yields the following scaling relation
\begin{equation}
  T_\mathrm{mg}(<r_\mathrm{ap}) \propto M(<r_\mathrm{ap})^{2/3} \, [\Delta \rho_x (z)]^{1/3}\,.
\end{equation}
Note the dependence on $\Delta \rho_x (z)$, which introduces a redshift dependence in the normalization of this scaling relation between cluster
temperature and mass.

Lastly, to compute the cylindrically-integrated SZ signal, $Y_{SZ}(<R_\mathrm{ap})$, we first calculate the Compton-$y$ parameter by integrating the thermal pressure of the gas along the line-of-sight using
\begin{equation}\label{eqn:compton}
    y_\mathrm{SZ}(R) = 2 \int_R^{r_\rmb} n_\rme(r) \frac{k_\rmB T_\rme(r)}{m_\rme c^2} \, \sigma_\rmT \, \frac{r \, \rmd r}{\sqrt{r^2 - R^2}}\,, 
\end{equation}
where $k_\rmB$, $m_\rme$, $c$, and $\sigma_\rmT$ are standard constants and $n_\rme (r)$ and $T_\rme (r)$ denote the electron gas number density and temperature profiles. The line-of-sight integration is performed out to $r_\rmb \equiv 2r_\mathrm{200m}$, which is roughly consistent with the radius of the accretion shock beyond which the pressure profile rapidly drops to the ambient pressure of the intergalactic medium \citep[see e.g.,][]{Molnar2009, Lau2015}. We then integrate $y_\mathrm{SZ}(R)$ over the aperture using
\begin{equation}
   Y_\mathrm{SZ}(< R_\mathrm{ap}) = 2 \pi \int_0^{R_\mathrm{ap}} y_\mathrm{SZ}(R) R \,\rmd R .
\end{equation}
As can be seen from equation~\eqref{eqn:compton}, the SZ signal is proportional to the product of the gas density and temperature. Hence, the self-similar \citet{Kaiser1986} model predicts that 
\begin{equation}
    \begin{split}
        Y_\mathrm{SZ}(<R_\mathrm{ap}) &\propto M_\mathrm{gas}(<r_\mathrm{ap}) \,  T_\mathrm{mg}(<r_\mathrm{ap}) \\ &\propto M(<r_\mathrm{ap})^{5/3} \, [\Delta \rho_x (z)]^{1/3}\,.
    \end{split}
\end{equation}

In Section \ref{ssec:omr}, we study the deviations of these observable-mass relations from self-similarity due to the injection of turbulence via mass assembly.  \new{The observable quantities and their self-similar scaling relations are summarized in Table \ref{tab:quantities}.}

\begin{table}
    \centering
    \begin{tabular}{c|c|c}
        Quantity & Mass Slope & Overdensity Slope \\
        \hline
        $Y_\mathrm{SZ}(<R_\mathrm{ap})$ & 5/3 & 1/3\\
        $M_\mathrm{gas}(<r_{\rm ap})$ & 1 & 0\\
        $T_\mathrm{mg}(<r_{\rm ap})$ & 2/3 & 1/3
    \end{tabular}
    \caption{\new{The observable cluster quantities studied in this work alongside their predicted power-law coefficients with respect to halo mass and spherical overdensity according to the self-similar model. The integrated Sunyaev-Zel'dovich signal, $Y_\mathrm{SZ}$, is calculated within a projected aperture, whereas the gas mass, $M_\mathrm{gas}$, and gas mass-weighted temperature, $T_\mathrm{mg}$, are computed within a spherical aperture.}}
    \label{tab:quantities}
\end{table}

\subsection{Mass accretion histories}\label{ssec:mah}

\new{We assume that the dark matter distribution of haloes follow the Navarro-Frenk-White (NFW) density profile \citep[][]{Navarro.etal.97} with enclosed mass 
\begin{equation}
    M(<r) = M_\mathrm{vir} \frac{f(c_\mathrm{vir} r / r_\mathrm{vir})}{f(c_\mathrm{vir})},
\end{equation}
where $M_\mathrm{vir}$, $r_\mathrm{vir}$, and $c_\mathrm{vir}$ are the halo virial mass, radius, and concentration,\footnote{$M_\mathrm{vir}$ is the mass enclosed within $r_\mathrm{vir}$, inside of which the mean density is equal to $\Delta_\mathrm{vir}(z)$ times the critical density. At $z=0$, $\Delta_\mathrm{vir}(z)\approx 100$, and is otherwise well-described by \citet{Bryan.Norman.98} for general $z$ and cosmology. The concentration is $c_\mathrm{vir} = r_\mathrm{vir} / r_\mathrm{s}$, with $r_\mathrm{s}$ the NFW scale radius.} respectively, and $f(x) = \ln(1+x) - x/(1+x)$.}

The mass assembly history, $M_\mathrm{vir}(z)$, tracks the \textit{main branch} of the halo, which is the branch of the halo merger tree that follows the main progenitor of the main progenitor of the main progenitor and so on \citep[halo merger trees are discussed in detail in Section 2.1 of][]{Jiang.vdBosch.16}. We compute \textit{individual} MAHs using the merger tree method described in \citet{Parkinson.etal.08}, a Monte Carlo approach based on the extended Press--Schechter \citep[EPS;][]{Bond.etal.91} formalism, which the method comparison project of \citet{Jiang.vdBosch.14} finds to perform the best at reproducing merger trees in simulations. 

\new{The EPS formalism gives the progenitor mass function (PMF), $n(M_\rmp, z_2 |  M_1, z_1) \, \rmd M_\rmp$, which specifies the average number of progenitor haloes with a mass of $M_\rmp \pm \rmd M_\rmp /2$ present at $z_2$ that merge into a descendant halo with a mass of $M_1$ at $z_1 < z_2$. Given a target halo mass of $M_\mathrm{vir,0}$ at redshift of observation $z_0$, one can sample a set of progenitor halo masses from the PMF, $M_{\rmp,1}$, $M_{\rmp,2}$, ..., $M_{\rmp,N}$, that at a previous time of $z_1 = z_0 + \Delta z$ satisfy $\sum_{i=1}^N M_{\rmp,i} = M_\mathrm{vir,0}$. Beginning at $z_0$, the merger tree is constructed by walking backwards in time with a temporal resolution of $\Delta z$ (which need not be constant along the tree), at each point sampling the progenitors of each descendant down to a mass resolution of $M_\mathrm{res}$. This $M_\mathrm{res}$ is typically a fixed fraction of the target halo mass, which we denote $\psi_\mathrm{res}=M_\mathrm{res}/M_\mathrm{vir,0}$; throughout this work, we use a mass resolution of $\psi_\mathrm{res}=10^{-4}$.} The \citet{Parkinson.etal.08} method generates merger trees based on the `binary method with accretion' of \citet{Cole2000} alongside a PMF modified from EPS theory to reproduce the merger statistics of the \textit{Millennium Simulation} \citep{Springel2005}. \new{For the construction of the tree, we use the timestep schedule motivated in \citet{Parkinson.etal.08}, which corresponds to $\Delta z \approx 10^{-3}$. However, as discussed in Section 2.2 of \citet{vdB2014}, for the purpose of computational efficiency, we down-sample the temporal resolution of the merger tree outputs to a timestep of $\Delta t = 0.1 t_\mathrm{ff}(z)$. The free-fall time for a halo with a critical overdensity of 200 at a redshift of $z$, $t_\mathrm{ff}(z)\propto (1+z)^{-3/2}$, is on the order of the halo dynamical time. Hence, there is little information added by using a smaller $\Delta t$; we have verified that our subsequent results are converged with respect to the merger tree timestep.}

These Monte Carlo MAHs are used in our analysis of the observable-mass relations in Sections~\ref{ssec:omr} and~\ref{ssec:mar}. When we are interested in the \textit{average} properties of a given halo of mass $M_\mathrm{vir,0}$ at redshift of observation $z_0$, we use the `universal model' of the average MAH described in \citet{vdB2014} (see their Section 4.1 and Appendix C). \new{In this case, we also trace the MAH back to the redshift that satisfies $\psi_\mathrm{res}=10^{-4} = M(z)/M_\mathrm{vir,0}$.} The average MAHs are used to study the properties of $f_\mathrm{nth}(r)$ and the HSE bias in Sections~\ref{ssec:fnth} and~\ref{ssec:hse}.

\begin{figure}
    \centering
    \includegraphics[width=0.45\textwidth]{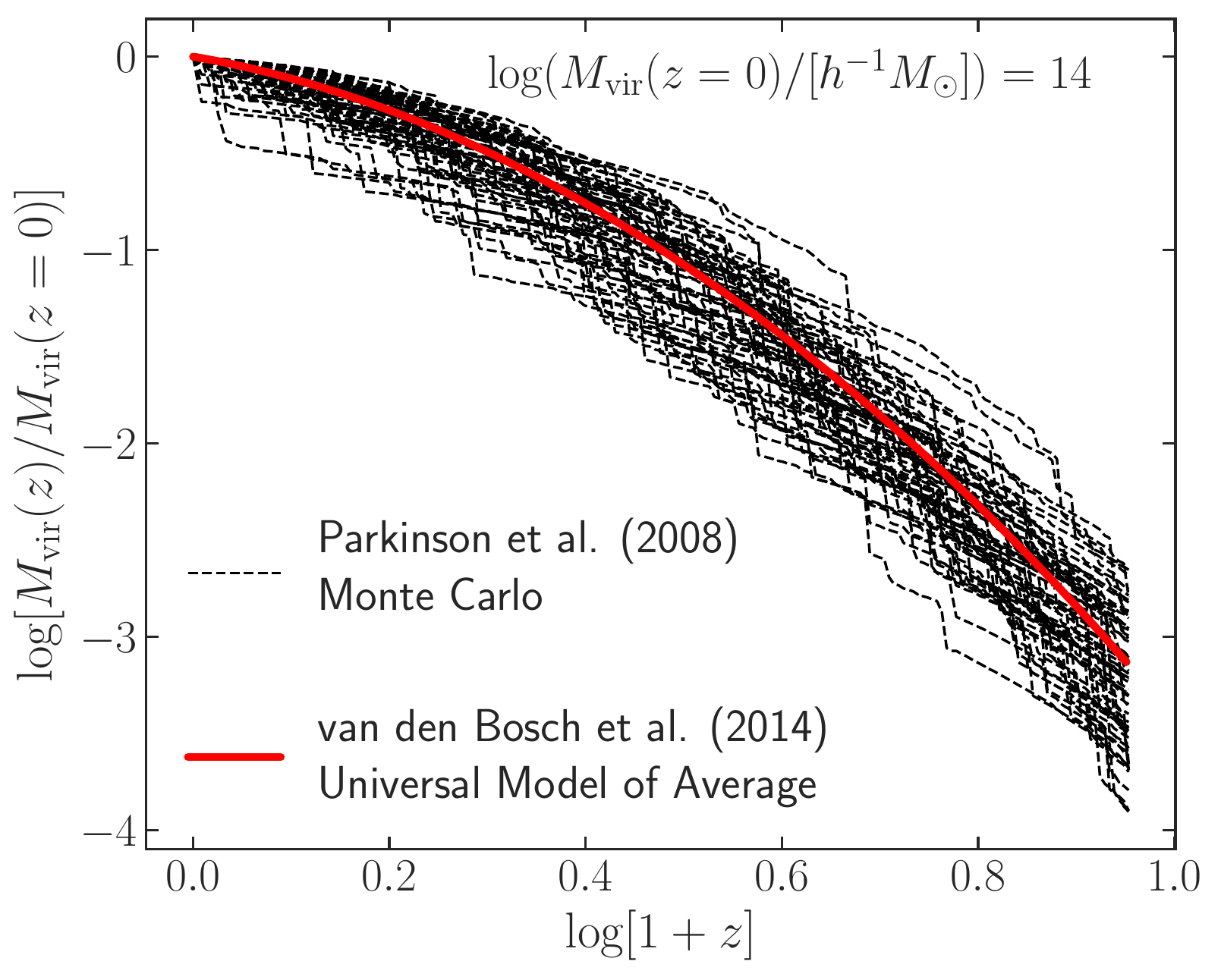}
    \caption{Example mass accretion histories, $M_\mathrm{vir}(z)$, for haloes with a final mass of $\log(M_\mathrm{vir}(z=0) / [h^{-1}M_\odot]) = 14$. The black, dashed lines correspond to individual MAHs generated by the Monte Carlo method of \citet{Parkinson.etal.08} whereas the thick, red line represents the `universal model' of the average MAH developed in \citet{vdB2014}. The individual MAHs are essential for studying the scatter in the observable-mass scaling relations (Sections~\ref{ssec:omr} and~\ref{ssec:mar}), whereas the average MAH is used for studying properties of the non-thermal pressure fractions and HSE mass bias (Sections~\ref{ssec:fnth} and~\ref{ssec:hse}).}
    \label{fig:mahs}
\end{figure}

For illustrative purposes, in Fig.~\ref{fig:mahs}, we show many different Monte Carlo-generated $M_\mathrm{vir}(z)$ trajectories for haloes with $\log(M_\mathrm{vir}(z=0) / [h^{-1}M_\odot]) = 14$. In addition, we overplot the average MAH predicted by the \citet{vdB2014} model for a halo of the same mass, demonstrating good agreement.

Concentrations are determined using the model of \citet{Zhao2009} as modified by \citet{vdB2014} to accurately reproduce the concentrations seen in \textit{Bolshoi}. The halo concentrations are given by
\begin{equation}\label{eqn:conc}
    c_\mathrm{vir} (M_\mathrm{vir}, t) = 4.0 \Bigg[ 1 + \bigg( \frac{t}{3.40t_{0.04}} \bigg)^{6.5} \Bigg]^{1/8}.
\end{equation}
\new{At proper time $t$, the halo has mass $M_\mathrm{vir}(t)$. The time $t_{0.04}$ is the proper time at which the halo's progenitor has accumulated a mass of $0.04M_\mathrm{vir}(t)$, which can be computed directly from the MAH. If $0.04M_\mathrm{vir}(t) < \psi_\mathrm{res}M_\mathrm{vir,0}$, we set $c_\mathrm{vir}(t) = 4$, which is the lower bound in the \citet{Zhao2009} model that all haloes tend toward at high $z$; we have verified that our results are insensitive to this choice. Note that each halo has a $c_\mathrm{vir}(t)$ trajectory determined solely by its MAH.} We have verified that the main results of this work are insensitive to the specific $c_\mathrm{vir}(M_\mathrm{vir},z)$ model used (we isolate the effect of the $c_\mathrm{vir}(M_\mathrm{vir},z)$ relation on our results in Section~\ref{ssec:omr}).

When discussing the effect of the mass assembly history on deviations from the self-similar observable-mass relations, it is convenient to use a summary statistic of $M_\mathrm{vir}(z)$ that captures the mass accretion rate (MAR) over a finite period of time. Throughout, we use the definition of the MAR introduced in \citet{Diemer2017a}, which encapsulates the change in $M_\mathrm{200m}$ over one dynamical (or crossing) time, $t_\mathrm{dyn} = 2 r_\mathrm{200m} / v_\mathrm{200m}$, where $v_\mathrm{200m}$ is the circular velocity at $r_\mathrm{200m}$. This MAR is written as
\begin{equation}\label{eqn:mar}
    \Gamma = \frac{\log[M_\mathrm{200m}(a_\mathrm{obs})] - \log[M_\mathrm{200m}(a_1)]}{\log(a_\mathrm{obs}) - \log(a_1)} ,
\end{equation}
where $a_\mathrm{obs}=(1+z)^{-1}$ corresponds to the redshift of observation $z$ and $a_1 = a(t_\mathrm{obs} - t_\mathrm{dyn})$ is the scale factor one dynamical time prior to observation. \new{In practice, the MAHs are discretely sampled in time, so we approximate $a_1$ and $M_\mathrm{200m}(a_1)$ as the value of the scale factor and mass at the timestep that is closest to $t(a_1)$ in the MAH output. Note that the halo concentration anti-correlates with $\Gamma$ --- more relaxed systems tend to be more highly concentrated.}

\subsection{Total pressure and gas density profiles}\label{ssec:presdens}

\new{Assuming the dark matter halo is well-described by the NFW density profile}, \citet[][hereafter \citetalias{Komatsu.Seljak2001}]{Komatsu.Seljak2001} develop a polytropic gas model for clusters in HSE where the thermal pressure is $P_\mathrm{th}(r) \propto \rho_\mathrm{gas}(r) T(r) \propto \rho_\mathrm{gas}(r)^\new{\gamma}$, with $\new{\new{\gamma}}$ the polytropic index \new{(note that this is different than the adiabatic coefficient)}. 

However, by studying simulated galaxy clusters, \citet{Shaw2010} find that a polytropic model describes the \textit{total} pressure profile, $P_\mathrm{tot}(r)$, better than $P_\mathrm{th}(r)$. \new{Specifically, \citet{Shaw2010} report that $P_\mathrm{tot}(r)$ follows a polytrope with fixed $\gamma=1.2$ over four decades in $\rho_\mathrm{gas}(r)$, indicating that $\gamma$ does not vary with cluster-centric radius.} Hence, we use the model of \citetalias{Komatsu.Seljak2001} to compute $P_\mathrm{tot}(r) \propto \rho_\mathrm{gas}(r) T_\mathrm{eff}(r) \propto \rho_\mathrm{gas}(r)^\new{\gamma}$. Here, $T_\mathrm{eff}(r)$ is an effective temperature profile that accounts for both the thermal and non-thermal pressure, which we write as $T_\mathrm{eff}(r) \equiv T_{\mathrm{eff},0} \theta(r)$. The resulting total pressure and gas density are thus parameterized as
\begin{equation}
    P_\mathrm{tot}(r) = P_0 \theta(r)^{\frac{\new{\gamma}}{\new{\gamma}-1}} \quad \textrm{and} \quad \rho_\mathrm{gas}(r) = \rho_0 \theta(r)^{\frac{1}{\new{\gamma}-1}},
\end{equation}
where all of $P_0$, $\rho_0$, and $\theta(r)$ depend on $M_\mathrm{vir}$ and $c_\mathrm{vir}$.\footnote{The native mass definition of the \citetalias{Komatsu.Seljak2001} gas model and our MAH models (Section~\ref{ssec:mah}) is that of $\Delta_\mathrm{vir}(z)$. We convert between mass and radius definitions using the concentration model in equation~\eqref{eqn:conc} and adopt the `200m' and `500c' mass conventions for comparisons with various simulation and observational results.} In addition, for reasons explained below, we have that $\new{\gamma}=\new{\gamma}(c_\mathrm{vir})$. Plugging these parameterizations into the HSE equation (equation~[\ref{eqn:hse}]), where we now use $P(r)=P_\mathrm{tot}(r)$, yields
\begin{equation}
    \theta(r, M_\mathrm{vir}, c_\mathrm{vir}) = 1 + \frac{\new{\gamma} - 1}{\new{\gamma}} \frac{\rho_0}{P_0} [\Phi(0) - \Phi(r)]\,,
\end{equation}
with $\Phi(r)$ the NFW gravitational potential profile, given by
\begin{equation}
    \Phi(r) = - \frac{G M_\mathrm{vir}}{r_\mathrm{vir}} \frac{c_\mathrm{vir}}{f(c_\mathrm{vir})} \frac{\ln(1+c_\mathrm{vir}r/r_\mathrm{vir})}{c_\mathrm{vir}r/r_\mathrm{vir}} .
\end{equation}

\new{A core assumption of \citetalias{Komatsu.Seljak2001} is that the gas density profile traces the dark matter density profile in the outer halo. Under this assumption, the normalization of the mass-temperature relation (or, when $P_\mathrm{tot} \neq P_\mathrm{th}$, the mass-effective-temperature relation; i.e., $P_0 / \rho_0 \propto T_{\mathrm{eff},0}$) is determined by asserting that the slope of the dark matter and gas density profiles are the same at some matching radius, $r_*$. In order for the gas profile to trace the dark matter profile over a large radial range ($r_\mathrm{vir}/2 < r < 2r_\mathrm{vir}$), their slopes must agree for a range of $r_*$. Since the value of $P_0 / \rho_0$ should not depend on the choice of $r_*$, the polytropic index, $\gamma$, is set such that $P_0 / \rho_0$ is independent of $r_*$ (in other words, $\gamma$ solves ${\rm d}(P_0 / \rho_0)/{\rm d} r_* = 0$).} Since the shape of the dark matter density profile depends on halo mass via the mass-concentration relation, both $\new{\gamma}$ and $P_0/\rho_0$ also depend on $c_\mathrm{vir}$. We follow \citetalias{Komatsu.Seljak2001}, adopting their polynomial fitting functions given by their equations~(25) and~(26). \new{Both $\gamma$ and $P_0/\rho_0$ grow with $c_\mathrm{vir}$ and thus tend to be lower in disturbed systems with high MAR.}

Motivated by the discussion of $M_\mathrm{gas}(<r_\mathrm{ap}) / M(<r_\mathrm{ap})$ in Section~\ref{ssec:selfsim}, we set the normalization of $\rho_\mathrm{gas}(r)$ such that $M_\mathrm{gas}(< 2 r_\mathrm{200m})$ is equal to the cosmic baryon fraction $\Omega_\mathrm{b} / \Omega_\mathrm{m}$ times $M(<2r_\mathrm{200m})$; our results are insensitive to the exact radius used to set this normalization.

\subsection{Non-thermal pressure profile}\label{ssec:nth}

In order to calculate cluster observables, we need to disentangle $P_\mathrm{th}(r)$ from $P_\mathrm{tot}(r) = P_\mathrm{th}(r) + P_\mathrm{nth}(r)$, where $P_\mathrm{nth}(r)$ is the non-thermal pressure. We determine the non-thermal pressure by following the analytical model of \citet[][hereafter \citetalias{Shi2014}]{Shi2014}. From the \citetalias{Komatsu.Seljak2001} total pressure and gas density profiles, we calculate the total velocity dispersion of the gas (per degree of freedom) as
\begin{equation}\label{eqn:sigma_tot}
    \new{\sigma_\mathrm{tot}^2(r, M_\mathrm{vir}, c_\mathrm{vir}) = P_\mathrm{tot}(r) / \rho_\mathrm{gas}(r) = (P_0 / \rho_0) \theta(r).}
\end{equation}
We emphasize that $\sigma_\mathrm{tot}^2$ evolves in time due to changing mass \textit{and} concentration, \new{i.e., $M_\mathrm{vir}(z)$ and $c_\mathrm{vir}(z)$}. The ansatz of \citetalias{Shi2014} is that the turbulent energy (per unit mass per degree of freedom), $\sigma_\mathrm{nth}^2 (r)$, dissipates on a timescale proportional to the eddy turn-over time of the largest eddies, which is in turn proportional to the local orbital time, $t_\mathrm{dis}(r) = \beta t_\mathrm{orb}(r) / 2$, and a fraction $\eta$ of the total energy injected into the cluster via mass growth is converted into turbulence. Based on this ansatz, the non-thermal energy evolves as
\begin{equation}\label{eqn:evolve}
    \frac{\rmd \sigma\nth^2}{\rmd t} = -\frac{\sigma\nth^2}{t_\mathrm{dis}} + \eta \frac{\rmd \sigma_\mathrm{tot}^2}{\rmd t} .
\end{equation}
The free parameters are calibrated against cosmological simulations in \citet{Shi2015} to $\beta = 1$ and $\eta=0.7$, which we adopt throughout. Determining $\sigma\nth^2$ at redshift $z$ is an initial value problem; \citetalias{Shi2014} find that the results are insensitive to the initial redshift, $z_i$, and initial $\sigma\nth^2$, \new{opting to use $z_i=6$ and $\sigma\nth^2(r,z_i) = \eta \sigma_\mathrm{tot}^2(r,z_i)$. Rather than begin at a fixed $z_i$, our initial redshift varies based on the $z_i$ that satisfies $M(z_i) = \psi_\mathrm{res} M_\mathrm{vir,0}$. For example, for $\psi_\mathrm{res} = 10^{-4}$, haloes in the mass range $12 \leq \log(M_\mathrm{vir}(z=0)/[h^{-1}M_\odot]) \leq 15.5$ have a $z_i$ that varies from $5-20$, with a distribution that peaks at $z_i=10$. Hence, our initial conditions use $\psi_\mathrm{res}=10^{-4}$ and $\sigma\nth^2(r,z_i) = \eta \sigma_\mathrm{tot}^2(r,z_i)$.}\footnote{We also impose the physical constraint that whenever $\sigma\nth^2(r, t_{j-1}) < -\rmd \sigma\nth^2(r, t_{j-1})$, then $\sigma\nth^2(r, t_{j})=0$ rather than becoming negative; this can happen occasionally near the halo center, where $t_\mathrm{dis}$ is small.} \new{We have verified that our subsequent results do not change if we decrease $\psi_\mathrm{res}$ (i.e., increase $z_i$); additionally, we have verified that at $\psi_\mathrm{res}=10^{-4}$, the results are insensitive to the initial $\sigma\nth^2(r,z_i)$ profile used. This is because at the corresponding} sufficiently high initial redshift, the time between $z_i$ and $z$ is many multiples of the initial $t_\mathrm{dis}(r)$ (i.e., $15-10^4$ times) for all $r$ and $z$ of interest, \new{dissipating away the} initial value of $\sigma\nth^2$. Note that $\sigma\nth^2$ is evolved independently for each cluster-centric radius $r$.

The timestep used in the $\sigma\nth^2$ evolution \new{is the same as that of the merger tree, which} corresponds to 10\% of the instantaneous cluster free-fall time at a critical overdensity of 200. At each timestep, the halo mass and concentration are updated based on the MAH model described \new{above} in Section \ref{ssec:mah}. The updated $M_\mathrm{vir}(z_j)$ and $c_\mathrm{vir}(z_j)$ result in updated $\new{\gamma}(c_\mathrm{vir})$, $(P_0/\rho_0)(c_\mathrm{vir})$, and $\Phi(r, M_\mathrm{vir}, c_\mathrm{vir})$, which we can use to compute $\sigma_\mathrm{tot}^2 (r, z_j)$. We then compute
\begin{equation}\label{eqn:evol}
    \frac{\rmd \sigma_\mathrm{tot}^2}{\rmd t} (r, z_j) = \frac{\sigma_\mathrm{tot}^2 (r, z_j) - \sigma_\mathrm{tot}^2 (r, z_{j-1})}{t(z_j) - t(z_{j-1})},
\end{equation}
after which we can compute $\rmd \sigma_\mathrm{nth}^2 / \rmd t$ to get our updated $\sigma\nth^2(r,z_j)$. Note that different $M_\mathrm{vir}(z)$ and $c_\mathrm{vir}(z)$ trajectories thus result in different $(\rmd \sigma_\mathrm{tot}^2 / \rmd t)(r,z)$ trajectories. Hence, for fixed $M_\mathrm{vir}$ at observation redshift $z$, \textit{all variance in the $\sigma_\mathrm{nth}^2$ profiles is due to inter-cluster differences in MAHs.} \new{We have checked our results for convergence with respect to timestep in the $\sigma\nth^2$ evolution, finding that the final $\sigma\nth^2$ change insignificantly when the size of the timestep is reduced by a factor of five.}

With $\sigma_\mathrm{nth}^2 (r)$ computed, we define the non-thermal energy fraction as $f\nth(r) = \sigma_\mathrm{nth}^2 (r) / \sigma_\mathrm{tot}^2 (r)$. The thermal pressure profile is then $P_\mathrm{th}(r) = [1-f\nth(r)]P_\mathrm{tot}(r)$.\footnote{Note that the temperature is related to the effective temperature (Section~\ref{ssec:presdens}) via $T(r) = [1-f_\mathrm{nth}(r)] T_\mathrm{eff}(r)$.} From $P_\mathrm{th}(r)$ and $\rho_\mathrm{gas}(r)$, we can compute the aforementioned cluster observables. In addition, we explore the mass and redshift dependence of the non-thermal pressure fraction and its implications for the HSE mass bias in Sections~\ref{ssec:fnth} and~\ref{ssec:hse}, respectively.

\new{We emphasize that} all results have been tested for convergence with respect to (i) the temporal resolution of the MAH and associated $\sigma\nth^2$ equation integration, \new{(ii) the initial conditions used for the integration of $\sigma\nth^2$ (i.e., $\sigma\nth^2(r,z_i)$ and $\psi_\mathrm{res}$),} (iii) the spatial resolution of the cluster profiles used to integrate the observables, and, where relevant, (iv) the number of MC-generated MAH realizations used to compute observable-mass relationships.

\subsection{Quantifying scaling relations}\label{ssec:quant}

\begin{figure}
    \centering
    \includegraphics[width=0.45\textwidth]{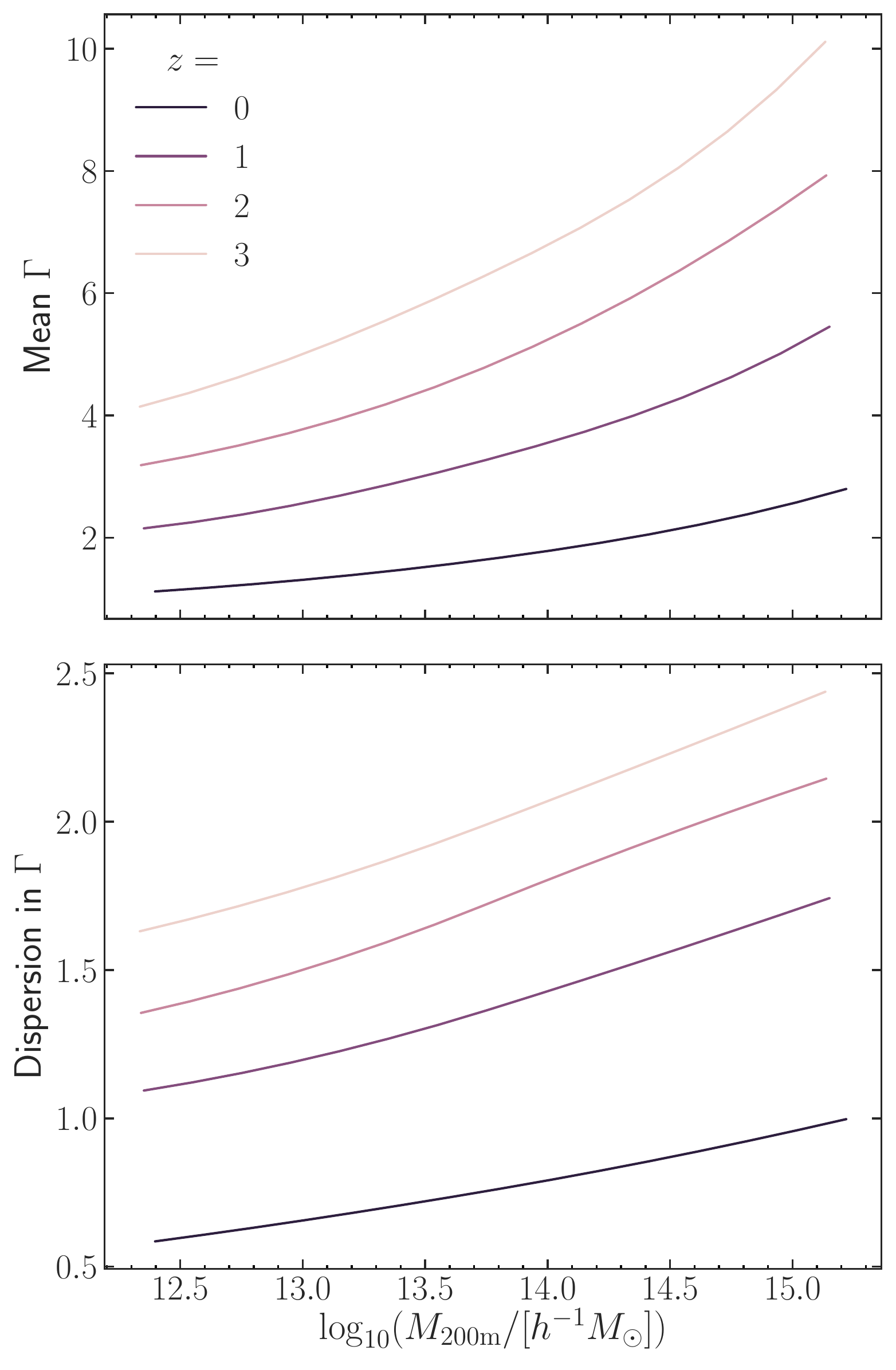}
    \caption{The mean of and dispersion in halo MARs, $\Gamma$ (defined in equation~[\ref{eqn:mar}]), as a function of halo mass and redshift for halo MAHs generated via the \citet{Parkinson.etal.08} method. The $\Gamma$ distribution is skewed (most strongly at low $M_\mathrm{200m}$ and $z$); hence, ``dispersion'' is defined as half of the $16-84$ percentile range. Note that both the mean of and dispersion in $\Gamma$ grow with $M_\mathrm{200m}$ and $z$. These trends are ultimately responsible for the same trends seen in the non-thermal pressure fractions (Section~\ref{ssec:fnth}) and for the increased scatter and decreased normalization in the scaling relations as $z$ increases (Section~\ref{ssec:omr}).}
    \label{fig:mar_mean_std}
\end{figure}

In our analysis of cluster scaling relations, we study individual, Monte Carlo-generated halo MAHs using the merger tree method of \citet{Parkinson.etal.08}. For each redshift of observation and cosmology considered, we generate {10,000} MAHs for haloes sampled uniformly in the mass range $12 \leq \log(M_\mathrm{vir}(z) / [h^{-1}M_\odot]) \leq 15.5$. \new{For consistency with other studies, our analysis uses the mass range of $14 \leq \log(M_\mathrm{200m}/[h^{-1}M_\odot]) \leq 15.6$ (a total of {${\sim}$4,500} clusters), but we use the lower-mass systems to check for any dependence on the mass cutoff in the scaling relations (as well as study how properties of $\Gamma$ depend on halo mass in Fig.~\ref{fig:mar_mean_std}). In the cluster mass regime}, our assumption of a mass-independent $M_\mathrm{gas}(<r_\mathrm{ap}) / M(<r_\mathrm{ap})$ ratio is well-justified (see the discussion in Sections~\ref{ssec:selfsim} and~\ref{ssec:presdens}). For each MAH, the $f_\mathrm{nth}(r)$ profile is evolved to the redshift of observation. Then, spanning a range of apertures, $r_\mathrm{ap}$, we compute the observables, $M_\mathrm{gas}(<r_\mathrm{ap})$, $T_\mathrm{mg}(<r_\mathrm{ap})$, and $Y_\mathrm{SZ}(<R_\mathrm{ap})$, and the corresponding halo mass, $M(<r_\mathrm{ap})$. We aim to elucidate how the slope, normalization, and scatter of the observable-mass relationships evolve with redshift and depend on aperture and cosmology. \textit{Note that we use the same aperture to calculate both the observable and the enclosed mass}. It is sometimes the case in observational studies that the mass is measured within one aperture (e.g., $r_\mathrm{500c}$) and the observable is measured within a larger aperture (e.g., $Y_\mathrm{SZ}[<5R_\mathrm{500c}]$), which can introduce additional effects due to the mass-concentration relation. We emphasize that in the limit that $f_\mathrm{nth}=0$, the observables are computed purely from the \citetalias{Komatsu.Seljak2001} model with $P_\mathrm{th}=P_\mathrm{tot}$, yielding the self-similar cluster scaling relations discussed in Section~\ref{ssec:selfsim} with no scatter or deviation aside from that due to the mass-concentration relationship; thus, all scatter is due to the variance in the halo MAHs and its impact on the $f_\mathrm{nth}$ profile and halo concentrations. In particular, increased $f_\mathrm{nth}$ will result in $T_\mathrm{mg}$ and $Y_\mathrm{SZ}$ decreasing and falling below the self-similar curve.

For each observable, $X_\mathrm{obs}(<r_\mathrm{ap})$, we compute the best-fit power-law relationship
\begin{equation}\label{eqn:omr}
    X_\mathrm{obs}(<r_\mathrm{ap}) = 10^\alpha \bigg(\frac{M(<r_\mathrm{ap})}{[10^{14} h^{-1} M_\odot]}\bigg)^\beta ,
\end{equation}
with $\alpha$ the normalization and $\beta$ the power-law slope. We then compute the \new{(natural)} logarithmic residuals as
\begin{equation}\label{eqn:resid}
    \new{\mathcal{R} = \ln(X_\mathrm{obs, true}) -\ln(X_\mathrm{obs, fit})},
\end{equation}
where $X_\mathrm{obs, fit}$ is computed from equation~\eqref{eqn:omr} given the $M(<r_\mathrm{ap})$ of each halo.\footnote{Note that our residual definition is opposite in sign to that which is normally used in the literature. As we show in Section~\ref{ssec:mar}, $\mathcal{R}$ as defined in equation~\eqref{eqn:resid} correlates with the halo MAR.}

We find that the \new{ln-residuals} for the $T_\mathrm{mg}-M$ and $Y_\mathrm{SZ}-M$ relations are \textit{not} normally distributed due to a strong \new{left}-skew (i.e., there is a long tail towards large, \new{negative} $\mathcal{R}$). As we illustrate in Section~\ref{ssec:mar}, this is directly due to a right-skew in the recent MARs of the haloes and a correlation between $\Gamma$ and non-thermal pressure support, which ultimately suppresses $Y_\mathrm{SZ}$. As shown in Fig.~\ref{fig:mar_mean_std}, the mean of and variance in $\Gamma$ grows with both halo mass and redshift for MAHs generated via the \citet{Parkinson.etal.08} method; this is directly responsible for a variety of trends in Section~\ref{sec:results}. Note that part of the strong right-skew is due to the fact that the MAR is bounded from below by zero, but is not bounded from above. A deviation from normality (and log-normality) of the residual distribution of $Y_\mathrm{SZ}-M$ in the form of a \new{left}-skew is also seen, albeit to a milder degree, in the non-radiative (NR) hydrodynamically-simulated clusters of \citet{Battaglia2012} (see their Fig. 22), indicating that in the absence of additional sources of non-thermal pressure beyond that introduced due to the halo assembly history, the residual distribution does indeed reflect the distribution of halo MARs. However, with the addition of radiative cooling, star formation, supernovae feedback, and AGN feedback, \citet{Battaglia2012} find that the residual distribution of $Y_\mathrm{SZ}-M$ approaches normality (\textit{not} log-normality). The relationship between the MAR and the $Y_\mathrm{SZ}-M$ residuals will be discussed in more detail in Section \ref{ssec:mar}.

The correlation between $\Gamma$ and non-thermal pressure support also causes the scatter in the scaling relations to increase systematically with halo mass. Regardless of the \new{ln-residual} distribution's deviation from normality and heteroscedasticity, ordinary least squares remains the best linear unbiased estimator of the mass-observable regression coefficients \citep{Plackett1950}. These details, however, do affect how we should report the scatter seen in the observable-mass relations. Typically, in analysis of both simulated and observed clusters, the residual distribution is assumed to be log-normal with mean zero. \new{Under this assumption, one can simply report the scatter as the standard deviation of the ln-residuals, $\sigma_{\mathcal{R}}$, which approximates the fractional/percent scatter, $\sigma_{\Delta X} / X$, to within 5\% (10\%) accuracy for $\sigma_{\mathcal{R}} \leq 0.1$ (0.2).} Due to the substantial deviation from log-normality in our case and in an effort to make comparisons to results in the literature, we report scatter instead based on half of the $16-84$ percentile range of $\mathcal{R}$. Our reported percent scatters are smaller by roughly $1\%$ (in absolute units, not relative) than they would be if we instead used the standard deviation of the \new{ln-residuals}.

\section{Results}\label{sec:results}

We start this section off by exploring the non-thermal pressure fraction profiles of the average cluster observed with a given mass at a particular redshift (Section~\ref{ssec:fnth}). We then study the resulting average HSE mass bias introduced due to non-thermal pressure support (and its dependence on halo mass and redshift) in Section~\ref{ssec:hse}. We proceed to calculate the scaling relations of samples of individual clusters, studying their dependence on aperture radius, cosmology, redshift, and halo mass limit in Section~\ref{ssec:omr}. Lastly, we identify a strong correlation between the halo MAR and the $Y_\mathrm{SZ}-M$ residual in Section~\ref{ssec:mar}, briefly discussing the potential utility of such a relationship.

\subsection{Non-thermal pressure fractions}\label{ssec:fnth}

\begin{figure*}
    \centering
    \includegraphics[width=\textwidth]{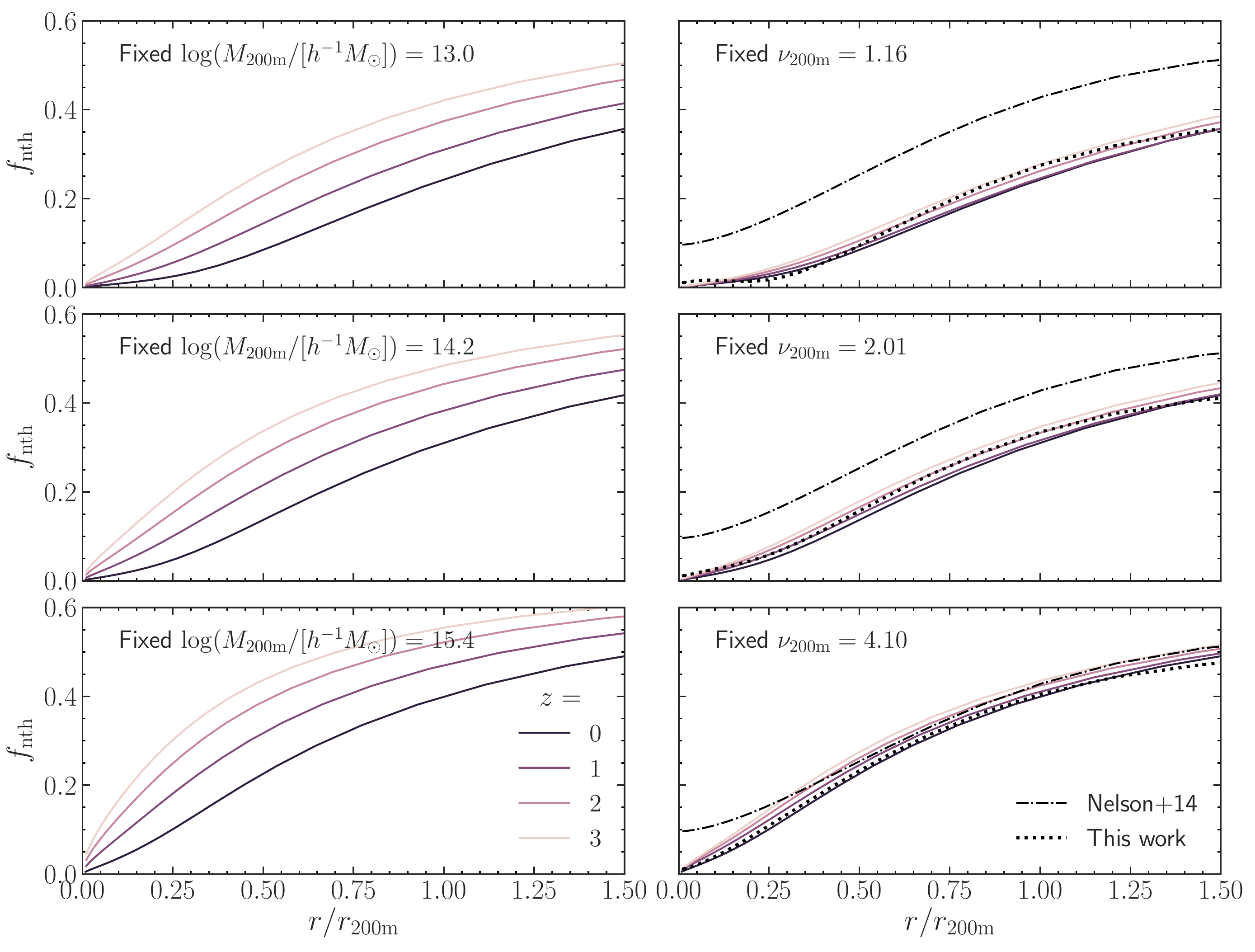}
    \caption{The non-thermal pressure fraction profiles of clusters, $f_\mathrm{nth} (r/r_\mathrm{200m})$, as predicted by the \citetalias{Shi2014} model combined with the `universal model' of the MAH from \citet{vdB2014}. (\textit{left}) Each panel holds the cluster mass, $M_\mathrm{200m}$, fixed and varies the redshift of observation. As either $M_\mathrm{200m}$ or $z$ increases, the non-thermal pressure fraction increases due to the increased recent mass accretion rate. (\textit{right}) Each panel holds the peak height, $\nu_\mathrm{200m}$, fixed such that the $z=0$ mass is the same as that in the corresponding left panel. \new{The masses corresponding to each peak height at the different redshifts are listed in Table~\ref{tab:nu_m}.} There is minimal redshift evolution in $f_\mathrm{nth} (r/r_\mathrm{200m})$ at fixed peak height. The `universal profile' seen in the simulated clusters of \citet{Nelson2014} is over-plot (dot-dashed line), illustrating the peak height-dependence that was not seen in their cluster sample due to their limited $z=0$ mass range. Our fitting function described by equation~\eqref{eqn:fnth_fit} and Table~\ref{tab:fit} (dotted line) incorporates $\nu_\mathrm{200m}$-dependence and reproduces the \citetalias{Shi2014} model at $z=1$ at roughly 10\% accuracy in the radial range of $0.2 \leq r/r_\mathrm{200m} \leq 2.0$.}
    \label{fig:fnth_vs_r}
\end{figure*}

Since the cornerstone of our analysis is the \citetalias{Shi2014} model of the non-thermal pressure, we first study its predicted $f_\mathrm{nth}(r)$ profiles for an average cluster of mass $M_\mathrm{200m}$ observed at $z$ using the `universal model' of the MAH from \citet{vdB2014}. In Fig.~\ref{fig:fnth_vs_r}, we plot the $f_\mathrm{nth}(r)$ profiles for clusters of several different masses as a function of $r/r_\mathrm{200m}$. The choice of $r_\mathrm{200m}$ is motivated by \citet{Nelson2014}, who find that the $f_\mathrm{nth}(r/r_\mathrm{200m})$ profiles of their sample of NR hydrodynamically-simulated galaxy clusters is universal throughout their time evolution (this will be discussed more below); this universality is absent when normalized by $r_\mathrm{200c}$.

In the left panels of Fig.~\ref{fig:fnth_vs_r}, we hold $M_\mathrm{200m}$ fixed and show how the $f_\mathrm{nth}$ radial profile changes with observation redshift. As halo mass increases, the non-thermal pressure fraction increases. This can be explained by the fact that higher mass haloes assemble at later times \citep[e.g.,][]{Lacey.Cole.93, vdBosch.02, Li2008}; hence, their recent MAR will be higher than that of lower mass haloes (cf. Fig.~\ref{fig:mar_mean_std}). More non-thermal energy has been recently injected into a system with a higher recent MAR, which results in a larger $f_\mathrm{nth}$. We also see that at fixed halo mass, $f_\mathrm{nth}$ is larger for clusters observed at higher redshift. This can be explained similarly to the previous point: in order for a halo to obtain a mass of $M$ by $z_1 > z_2$, it must have accreted mass more rapidly than a halo with a mass of $M$ at $z_2$ (cf. Fig.~\ref{fig:mar_mean_std}). Note that the fraction of non-thermal pressure is substantial, especially in the cluster outskirts --- $f_\mathrm{nth}$ surpasses 50\% by around ${\sim}r_\mathrm{200m}$ for high-mass haloes and haloes at large $z$.

\begin{table}
    \centering
    \begin{tabular}{c|c|c|c|c}
        $z$ & 0 & 1 & 2 & 3 \\
        \hline
        $\nu_{\rm 200m}$ & \multicolumn{4}{c}{$\log(M_{\rm 200m}/[h^{-1}M_{\odot}])$} \\
        \hline
        1.16 & 13.00 & 11.61 & 10.30 & 9.11 \\
        2.01 & 14.20 & 13.13 & 12.14 & 11.27 \\
        4.10 & 15.40 & 14.60 & 13.88 & 13.27
    \end{tabular}
    \caption{\new{The peak heights studied in Fig.~\ref{fig:fnth_vs_r} and the corresponding halo masses at each redshift.}}
    \label{tab:nu_m}
\end{table}

We now explore the dependence of $f_\mathrm{nth} (r/r_\mathrm{200m})$ on peak height, $\nu_\mathrm{200m} = \delta_\rmc(z) / \sigma(M_\mathrm{200m})$.\footnote{Here, $\delta_\rmc(z) = \delta_\rmc(z=0) / D_+(z)$ is the critical overdensity for collapse \citep{Gunn1972}, $D_+(z)$ is the linear growth factor normalized to unity at $z=0$, and $\sigma(M_\mathrm{200m})$ is the RMS mass fluctuation in a Lagrangian volume corresponding to $M_\mathrm{200m}$.} In the right panels of Fig.~\ref{fig:fnth_vs_r}, we hold $\nu_\mathrm{200m}$ fixed to several different values and show how $f_\mathrm{nth}(r/r_\mathrm{200m})$ evolves with redshift in each case (i.e., $M_\mathrm{200m}$ is varied with $z$ such that $\nu_\mathrm{200m}$ remains constant). As is apparent, there is far less redshift evolution at fixed peak height than at fixed mass. We overplot the fitting function of \citet{Nelson2014} for their universal $f_\mathrm{nth}(r/r_\mathrm{200m})$ profile, finding that for the peak height consistent with the $z=0$ cluster masses studied in their work ($\nu_\mathrm{200m} \approx 4$), the predictions of the \citetalias{Shi2014} model agree well with what is seen in the simulations. There is an exception to this agreement, however, in the central regions of the clusters, where the model underpredicts the non-thermal pressure fraction compared to that seen in the \citet{Nelson2014} simulations. As discussed in \citet{Shi2015}, this is likely due to (i) the model's assumption of a one-to-one relationship between the cluster radius and the turbulence dissipation timescale (note that this assumption is the primary source of the $f_\mathrm{nth}$ radial dependence) and (ii) the potential need to incorporate radius and redshift dependence into $\eta$ to properly model the relative importance of high-Mach accretion shocks and low-Mach internal shocks. \new{Recently, \citet{Shi2018} found that the turbulence dissipation timescale measured in simulations is indeed shorter at smaller cluster radii and suggested that this is the case due to the stronger density stratification in the cluster core. Following this, \citet{Shi2019} confirmed the role of the density stratification and indicated that the buoyancy time ($t_\mathrm{BV}(r)$; i.e., the inverse of the Brunt-V{\" a}is{\" a}l{\" a} frequency) may more accurately capture the timescale of turbulence dissipation. The buoyancy time is nearly the same as the orbital time outside of the cluster core; however, for non-cool-core clusters, the core in the cluster entropy profile results in a core in the buoyancy time profile (i.e., $t_\mathrm{BV}$ approaches a constant, non-zero value as $r\rightarrow 0$). We find that using $t_\mathrm{BV}$ for the turbulence dissipation timescale causes the $f_\mathrm{nth}$ profiles to change by less than $10\%$ outside of the cluster core region (${\lesssim}0.2r_\mathrm{200m}$) relative to the fiducial case of $t_\mathrm{orb}$ --- this propagates to a difference of only a few percent in our subsequent $Y_\mathrm{SZ}$ model predictions, since the signal is dominated by the cluster outskirts.}

The model does not predict a universal $f_\mathrm{nth}(r/r_\mathrm{200m})$ profile, which clearly has a dependence on $\nu_\mathrm{200m}$ that, to good approximation, accounts for the dependence on both $M_\mathrm{200m}$ and $z$. At first, this appears to be at odds with the simulated clusters studied in \citet{Nelson2014}. However, they studied the evolution of a cluster sample through \textit{time}, with $M_\mathrm{200m}$ only spanning half an order of magnitude in the range $14.8 < \log(M_\mathrm{200m} / [h^{-1}M_\odot]) < 15.4$ at $z=0$. This detail, combined with their use of a $z$-dependent mass cutoff for the cluster sample, likely resulted in the \citet{Nelson2014} sample spanning an insufficient range in $\nu_\mathrm{200m}$ to isolate evolution in redshift from universality in $f_\mathrm{nth}(r/r_\mathrm{200m})$ at fixed $\nu_\mathrm{200m}$. 

Motivated by our finding that, to good approximation, $f_\mathrm{nth} = f_\mathrm{nth}(r/r_\mathrm{200m} \vert \nu_\mathrm{200m})$, we present a fitting function for the non-thermal pressure fraction that includes this $\nu_\mathrm{200m}$-dependence:
\begin{equation}\label{eqn:fnth_fit}
    f_\mathrm{nth}(\tilde{r} \vert \nu) = 1-\Big[A\big(1+e^{-(\tilde{r}/B)^C}\big)  \Big(\frac{\nu}{4.1}\Big)^{\frac{D}{(1+[\tilde{r}/E]^F)}}\Big] .
\end{equation}
Here, $\nu = \nu_\mathrm{200m}$ and $\tilde{r} = r / r_\mathrm{200m}$. The parameters of this function are calibrated to match the $z=1$ predictions of the model across $ 1.0 \leq \nu_\mathrm{200m} \leq 4.2$ and are listed in Table~\ref{tab:fit}. The fit, shown as dotted curves in Fig.~\ref{fig:fnth_vs_r}, is accurate to roughly 10\% over the radial range of $0.2 \leq r/r_\mathrm{200m} \leq 2.0$. \new{In a future work, we will further explore this $\nu_\mathrm{200m}$-dependence and the sensitivity of the $ f_\mathrm{nth}(r/r_\mathrm{200m})$ predictions to cosmology and more realistic definitions of the turbulence dissipation timescale.}

\begin{table} %[h!]
\centering
\begin{tabular}{lS[table-format=-1.3]}
\hline
Parameter  & \multicolumn1c{Value} \\
\hline
$A$ & 0.495 \\
$B$ & 0.719 \\ 
$C$ & 1.417 \\ 
$D$ & -0.166 \\ 
$E$ & 0.265 \\ 
$F$ & -2.116 \\
\hline
\end{tabular}
\caption{Calibrated parameters of the $f_\mathrm{nth} (r/r_\mathrm{200m}, \nu_\mathrm{200m})$ fitting function described by equation~\eqref{eqn:fnth_fit}, which reproduces the model non-thermal pressure fractions to roughly 10\% accuracy in the radial range of $0.2 \leq r/r_\mathrm{200m} \leq 2.0$ at $z=1$. Note that there is only a weak redshift dependence in the model predictions, as can be seen in Fig.~\ref{fig:fnth_vs_r}, so this fitting function can be easily used to make rough predictions regardless of redshift.}\label{tab:fit}
\end{table}

\subsection{Hydrostatic mass bias}\label{ssec:hse}

As discussed in the introduction, cluster mass inferences based on X-ray and SZ observations are typically made under the assumption of HSE between the observed thermal pressure profile and the gravitational potential. The true cluster mass, however, is related to the \textit{total} pressure profile, and thus any unaccounted-for sources of non-thermal pressure result in underprediction of the cluster mass. The $f_\mathrm{nth}$ profiles predicted by the \citetalias{Shi2014} model can be used to estimate the corresponding HSE mass bias.

\begin{figure}
    \centering
    \includegraphics[width=0.45\textwidth]{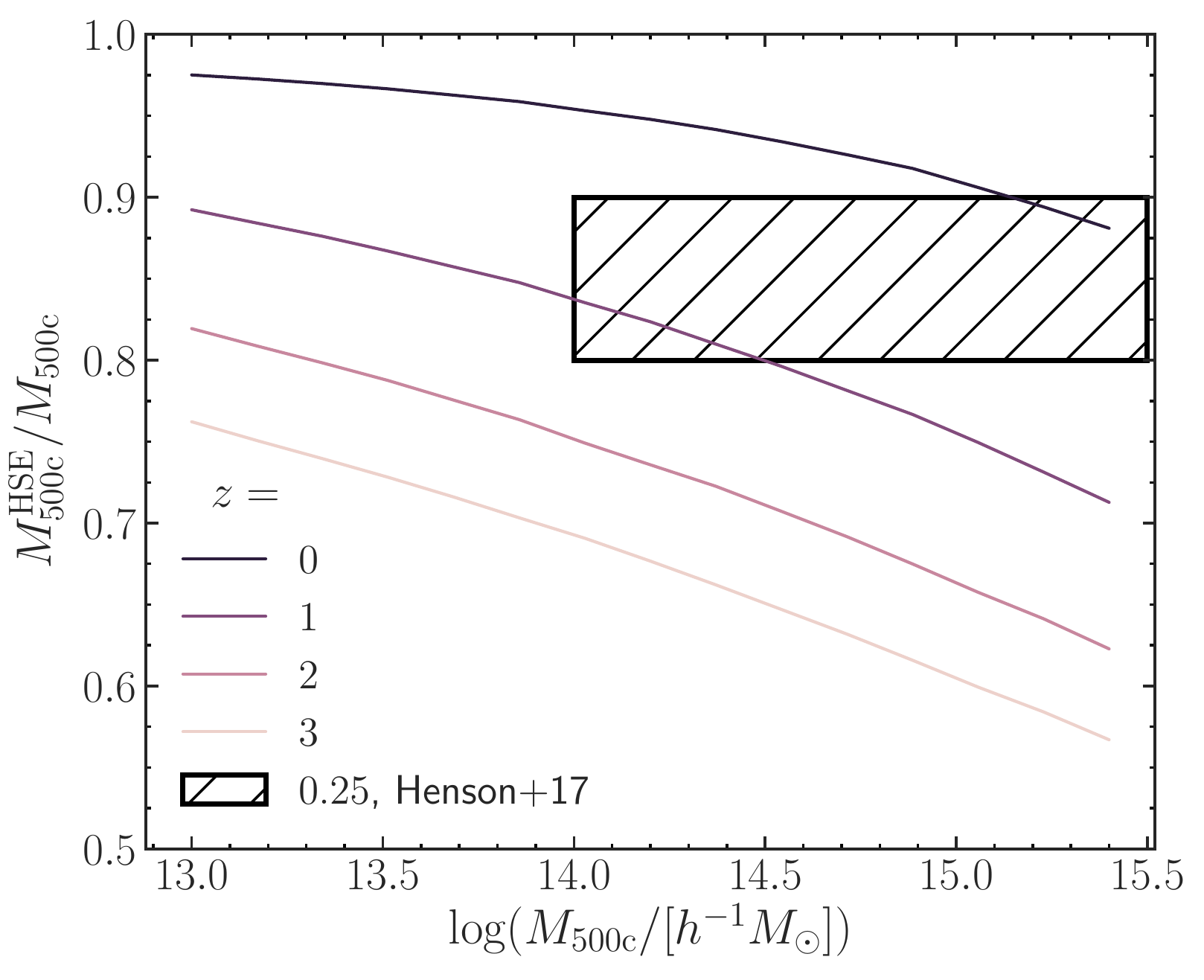}
    \caption{The HSE mass bias for $M_\mathrm{500c}$, computed using equation~\eqref{eqn:hseb}, as predicted using the \citetalias{Shi2014} model of the non-thermal pressure. The ``bias'' increases as $M^\mathrm{HSE}_\mathrm{500c} / M_\mathrm{500c}$ decreases. These results are roughly consistent with the simulated clusters studied in \citet[][at $z=0.25$, hatched black box]{Henson2017}, although our results can be considered lower bounds, as sources of non-thermal pressure in addition to those due to mass assembly are not modeled and will increase the bias. The mass bias increases substantially with redshift, motivating further simulation studies focused on the HSE bias redshift evolution.}
    \label{fig:hse_bias}
\end{figure}

From the HSE equation (i.e., equation~[\ref{eqn:hse}]), one can compute how much the true mass, $M$, is underpredicted ($M^\mathrm{HSE}$) as a function of mass and redshift. Assuming an accurate determination of the gas density and thermal pressure profiles, which can be made possible through the combination of X-ray and SZ observations \citep[e.g.,][]{Ameglio2009, Eckert2019, Ettori2019}, this underprediction is written as 
\begin{equation}\label{eqn:hseb}
    \frac{M^\mathrm{HSE}(<r)}{M(<r)} = \frac{\rmd P_\mathrm{th} / \rmd r}{\rmd P_\mathrm{tot} / \rmd r} = [1-f_\mathrm{nth}(r)] - P_\mathrm{tot}(r) \frac{\rmd f_\mathrm{nth} / \rmd r}{\rmd P_\mathrm{tot} / \rmd r}.
\end{equation}
Since $\rmd P_\mathrm{tot} / \rmd r$ is negative and $\rmd f_\mathrm{nth} / \rmd r$ is positive, this ratio should always be larger than $1-f_\mathrm{nth}(r)$ for measurements of mass enclosed within $r$. Note that this estimate of the HSE bias neglects potential effects due to the deviation from spherical symmetry and projection effects. In Fig.~\ref{fig:hse_bias}, we plot the predictions for $M^\mathrm{HSE}_\mathrm{500c} / M_\mathrm{500c}$ as a function of $M_\mathrm{500c}$ and redshift of observation. We use $r_\mathrm{ap} = r_\mathrm{500c}$ ($\approx 0.4 r_\mathrm{200m}$) since this is the aperture most commonly used for X-ray-based cluster mass estimation. At this radius, the \citetalias{Shi2014} model is in good agreement with the $f_\mathrm{nth}$ profiles of the simulated clusters of \citet{Nelson2014}, which only include NR hydrodynamics. Hence, additional sources of non-thermal pressure due to magnetic fields, cosmic rays, supernova feedback, among others, are not included and thus, we expect these estimates of the magnitude of the HSE bias to be \textit{lower bounds}. The \citetalias{Shi2014} model predicts that the magnitude of the HSE bias increases considerably with cluster mass and observation redshift. At $z=0$, HSE-based masses underestimate the true masses by less than $10\%$ even for the highest mass clusters. However, at $z\sim 2-3$, the HSE bias results in substantial underprediction of the true mass, by roughly $20\%$ at group scales and as much as $30-40\%$ for high-mass clusters.

In the $z=0.25$ simulated cluster sample from the \textit{BAHAMAS} \citep{McCarthy2017} and \textit{MACSIS} \citep{Barnes2017a} hydrodynamic simulations studied in \citet{Henson2017}, which include star formation, radiative cooling, and feedback from supernovae and AGN (hereafter referred to as ``full-physics'' simulations), the HSE bias found is $M^\mathrm{HSE}_\mathrm{500c} / M_\mathrm{500c}\approx 0.8- 0.9$ in the mass range $14 \leq \log(M_\mathrm{500c}/[h^{-1}M_\odot]) \leq 15.5$. Their bias is only marginally larger than that predicted by our model (Fig.~\ref{fig:hse_bias}), which is most likely due to the additional sources of non-thermal pressure captured in the full-physics simulations. Similarly, \citet{Ansarifard2020} reports a median of $M^\mathrm{HSE}_\mathrm{500c} / M_\mathrm{500c}\approx 0.9$ for simulated clusters at $z=0$ in a similar mass range. On the other hand, using synthetic X-ray observations, \citet{Barnes2020} report biases as significant as $M^\mathrm{HSE}_\mathrm{500c} / M_\mathrm{500c}\approx 0.7$ for simulated clusters at $z=0.1$ with $\log(M_\mathrm{500c}/M_\odot) \approx 15.3$ --- they find that this is primarily due to the use of a single temperature fit to the full cluster spectrum. \new{Additionally, \citet{Hurier2018} report a similar bias of $0.73\pm 0.07$ when using CMB lensing to estimate cluster masses, although they find no significant dependence on mass or redshift.} For reference, a larger bias is necessary ($M^\mathrm{HSE}_\mathrm{500c} / M_\mathrm{500c}\approx 0.6$) in order to resolve the tension between cosmological parameter estimates based on the cluster mass function and cosmic microwave background approaches \citep{Salvati2019}.

We emphasize that these calculations are based on the average MAH for a cluster observed with a given mass and redshift. Clusters that are more disturbed (i.e., have a higher recent MAR) will generally have larger biases than the average, as their non-thermal pressure fraction will be larger. In fact, the difference between the HSE bias of an individual cluster and the average (at fixed halo mass and redshift) should correlate with the MAR; as we discuss in Section~\ref{ssec:mar}, a strong correlation also arises between the residuals of the $Y_\mathrm{SZ}-M$ relation and the MAR. To date, we are not aware of any simulation studies that characterize the evolution of the HSE bias over a reasonably large range of redshifts. Based on the results of Fig.~\ref{fig:hse_bias}, such a study is warranted, as the redshift dependence of the HSE bias predicted will be important to account for in cluster count analyses that include high-$z$ cluster samples from future surveys.

\subsection{Cluster scaling relations}\label{ssec:omr}

\begin{figure*}
    \centering
    \includegraphics[width=\textwidth]{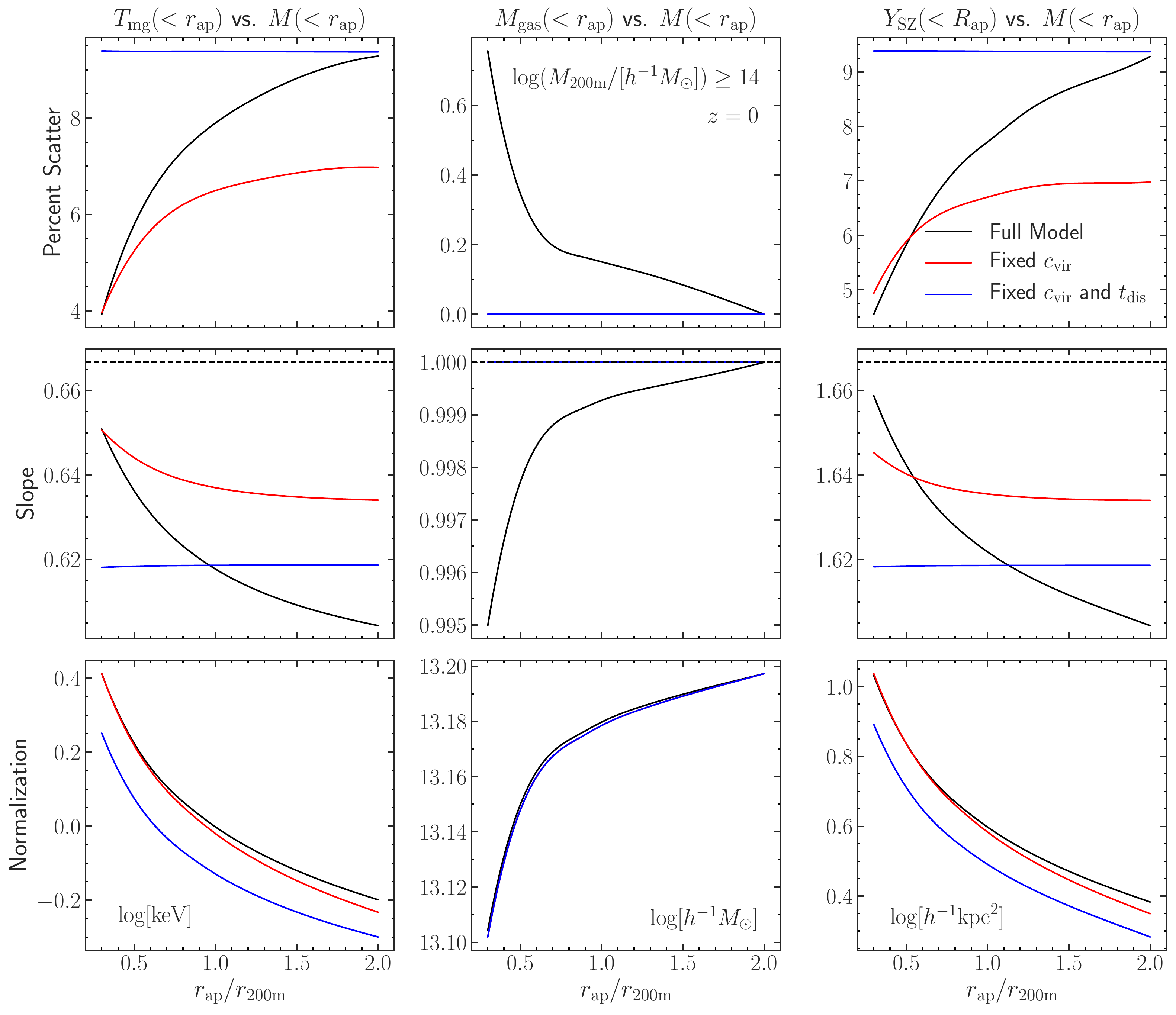}
    \caption{The best fit normalization, slope, and percent scatter \new{(i.e., half of the $16-84$ percentile interval of $\mathcal{R}$)} of the $z=0$ observable-mass relations, described by equation~\eqref{eqn:omr} for $T_\mathrm{mg}(<r_\mathrm{rap}) -M(<r_\mathrm{ap})$, $M_\mathrm{gas} (<r_\mathrm{rap}) -M(<r_\mathrm{ap})$, and $Y_\mathrm{SZ} (<R_\mathrm{ap})- M(<r_\mathrm{ap})$. The fit parameters are shown as a function of the aperture radius, $r_\mathrm{ap}$, in units of $r_\mathrm{200m}$; note that the same aperture is used to compute both the observable and the total mass. In these fits, {${\sim}$4,500} clusters in the mass range $14 \leq \log(M_\mathrm{200m}/[h^{-1}M_\odot]) \leq 15.6$ are used. The black dashed lines indicate the slopes predicted by the self-similar relations. The observables are computed using the ``full model'' described in Sections~\ref{ssec:presdens} and~\ref{ssec:nth} (black curves) as well as two simplified models, one of which holds $c_\mathrm{vir}=5$ fixed (red curves) in order to isolate the effects of the mass-concentration relation and another that holds both $c_\mathrm{vir}=5$ fixed \textit{and} replaces the radius-dependent turbulence dissipation timescale with its value at $r_\mathrm{200m}$ (blue curves), isolating the interaction between $r_\mathrm{ap}$ and the radius-dependence of $f_\mathrm{nth}(r)$. See the main text in Section~\ref{ssec:omr} for detailed explanations of the trends with $r_\mathrm{ap}$.
    }
    \label{fig:mor_complexity}
\end{figure*}

Having demonstrated that the average $f_\mathrm{nth}$ profiles (and resultant HSE biases) predicted by the \citetalias{Shi2014} model are in good agreement with predictions from hydrodynamical simulations, we proceed to use the model to study the cluster scaling relationships. In Fig.~\ref{fig:mor_complexity}, we plot the best fit normalization, slope, and percent scatter for the $z=0$ relations as a function of $r_\mathrm{ap}$. In order to provide insight into the model predictions and disentangle the nonlinear interactions between its various components, we calculate the cluster observables in three different ways. First, we compute cluster observables using the ``full model'' described in Sections~\ref{ssec:presdens} and~\ref{ssec:nth}. We then repeat the calculations while holding the halo concentrations fixed to $c_\mathrm{vir}=5$ (referred to as the ``fixed $c_\mathrm{vir}$ model''), isolating the impact of the mass-concentration relation. Going further, we perform a third set of calculations: while continuing to hold $c_\mathrm{vir}$ fixed, we now also replace the radius-dependent turbulence dissipation timescale with its value at $r_\mathrm{200m}$ (i.e., $t_\mathrm{dis}(r) = t_\mathrm{dis}(r_\mathrm{200m})$; referred to as the ``fixed $c_\mathrm{vir}$ and $t_\mathrm{dis}$ model''). This elucidates the impact of the radial dependence of $t_\mathrm{dis}(r)$. Note that in this final case, the $f_\mathrm{nth}$ profile is nearly constant with radius for a given halo and all variation in $f_\mathrm{nth}$ between haloes is due to variation in MAHs.

\subsubsection{$M_\mathrm{gas}-M$ relation}

Beginning with the $M_\mathrm{gas}-M$ relation (middle column of Fig.~\ref{fig:mor_complexity}), our model predicts no scatter in the absence of a MAH-dependence on the concentration. This is simply due to our use of the \citetalias{Komatsu.Seljak2001} model for $\rho_\mathrm{gas}$, which has no dependence on the halo MAH or $f_\mathrm{nth}$ but only on $c_\mathrm{vir}$ (the implications of this are discussed in more detail in Sections~\ref{sssec:ym} and~\ref{sssec:scatter}). In the full model, the scatter goes to zero and the slope goes to unity as $r_\mathrm{ap}\rightarrow 2r_\mathrm{200m}$; this is simply due to our chosen gas density normalization that $M_\mathrm{gas}(< 2 r_\mathrm{200m}) = (\Omega_\rmb / \Omega_\rmm) M(< 2 r_\mathrm{200m})$. The effect of the mass-concentration relation and its intrinsic scatter on the shape of both the dark matter and gas density profiles is responsible for the small change in slope (and increase in scatter) of $M_\mathrm{gas}-M$ as aperture radius decreases \citep[see e.g.,][]{Fujita2019}. Even with fixed concentrations, the difference between the gas and dark matter density profile shapes is responsible for a slight dependence on $r_\mathrm{ap}$ in the $M_\mathrm{gas}-M$ normalization.

\subsubsection{$T_\mathrm{mg}-M$ relation}

Next, we direct our attention to the mass-weighted temperature, $T_\mathrm{mg}$ (left-hand column of Fig.~\ref{fig:mor_complexity}). In the fixed $c_\mathrm{vir}$ and $t_\mathrm{dis}$ model, the scatter and slope are independent of aperture. More-massive clusters have larger MAR, which drives larger $f_\mathrm{nth}$ (see the left panels of Fig.~\ref{fig:fnth_vs_r}). For $f_\mathrm{nth}$ independent of radius, we have that $T_\mathrm{mg} \propto (1-f_\mathrm{nth})$, which, combined with the fact that $f_\mathrm{nth}$ grows with halo mass, results in the slope of $T_\mathrm{mg}-M$ lying below that of self-similarity. In this simplified model, the aperture-independent scatter in $T_\mathrm{mg}-M$ is also driven solely by variation in halo MAHs and is most sensitive to the mass evolution over the previous dynamical time. Moving on to the fixed $c_\mathrm{vir}$ model, we notice that incorporation of a radius-dependent $t_\mathrm{dis}(r)$ introduces a dependence on aperture into the slope and scatter of $T_\mathrm{mg}-M$. Higher mass clusters tend to have $f_\mathrm{nth}(r)$ profiles that are overall larger in magnitude \textit{and} grow more rapidly with radius (most notably in the inner radii; see once again the left panels of Fig.~\ref{fig:fnth_vs_r}). Hence, their temperature profiles will be more suppressed overall relative to self-similarity. In addition, since $\rmd f_\mathrm{nth}(r) / \rmd r$ increases with halo mass (in the inner radii), the slope of $T_\mathrm{mg}-M$ decreases further from self-similarity as $r_\mathrm{ap}$ increases. The scatter in $f_\mathrm{nth}(r)$ grows with radius due to the radially increasing $t_\mathrm{dis}(r)$; because of this, the scatter in $T_\mathrm{mg}-M$ grows with aperture radius. Lastly, by looking at the full model, we see two effects due to the mass-concentration relation. First, $c_\mathrm{vir}(M,z)$ results in further reduction in the $T_\mathrm{mg}-M$ slope away from self-similarity. Additionally, the variance in $c_\mathrm{vir}(M,z)$ propagates to additional scatter in $T_\mathrm{mg}-M$ that becomes more substantial as $r_\mathrm{ap}$ increases. Finally, the normalization of the $T_\mathrm{mg}-M$ relation decreases with increasing aperture for a simple reason. Since the cluster temperature decreases with radius, the mass-weighted temperature must decrease as the aperture radius increases. In addition, the pivot mass used for the relations is $M(<r_\mathrm{ap})=10^{14} h^{-1}M_\odot$ regardless of aperture. Hence, this pivot mass at larger $r_\mathrm{ap}$ corresponds to a smaller total (virial) mass and thus a lower temperature normalization.

\subsubsection{$Y_\mathrm{SZ}-M$ relation}
\label{sssec:ym}

Lastly, we turn to the integrated SZ signal, $Y_\mathrm{SZ}$ (right-hand column of Fig.~\ref{fig:mor_complexity}). Since $Y_\mathrm{SZ}$ is simply the cylindrically-integrated pressure profile, to good approximation $Y_\mathrm{SZ} \propto M_\mathrm{gas} T_\mathrm{mg}$. This relationship bares out straightforwardly in Fig.~\ref{fig:mor_complexity}, as the slope of the $Y_\mathrm{SZ}-M$ relation evolves roughly as the sum of the slopes of the $T_\mathrm{mg}-M$ and $M_\mathrm{gas}-M$ relations. \new{For the smallest values of $r_\mathrm{ap}$, we have verified that the slight disagreement between the slope of $Y_\mathrm{SZ}-M$ and the sum of the $T_\mathrm{mg}-M$ and $M_\mathrm{gas}-M$ slopes is simply due to projection effects that manifest due to the different impact of halo concentrations on spherically- and cylindrically-integrated quantities. The clusters from the NR hydrodynamics simulations of \citet{Stanek2010} yield a $Y_\mathrm{SZ}(<r_\mathrm{200c})-M_\mathrm{200c}$ relation slope of $1.651 \pm 0.003$. For comparison, and noting that $r_\mathrm{200c}\approx 0.6 r_\mathrm{200m}$, we find a slope in $Y_\mathrm{SZ}(<R_\mathrm{200c})-M_\mathrm{200c}$ of roughly 1.635 (note that this reduces slightly to 1.63 if we instead compute $Y_\mathrm{SZ}(<r_\mathrm{200c})-M_\mathrm{200c}$ with a spherically-integrated $Y_\mathrm{SZ}$, which is not shown).} In our calculations, the only source of scatter in $M_\mathrm{gas}-M$ is the mass-concentration relation. However, as described above, the variance in the cluster MAHs drives the scatter in $T_\mathrm{mg}-M$ and increases considerably with aperture. Thus, in our model, the scatter in the $Y_\mathrm{SZ}-M$ relation is driven predominantly by the scatter in the $T_\mathrm{mg}-M$ relation. \final{A more realistic model of the gas density profile that incorporates additional baryonic processes will introduce additional variance into $M_\mathrm{gas}-M$, as well as stronger covariance between $M_\mathrm{gas}$ and $T_\mathrm{mg}$ \citep[see e.g.,][]{Stanek2010}, which will ultimately increase the scatter in $Y_\mathrm{SZ}-M$. Hence, our scatter estimates should be regarded as lower bounds (see additional discussion in Section~\ref{sssec:scatter}).} Regarding the reduction in normalization and slope of $Y_\mathrm{SZ}-M$ with increasing aperture, we find similar trends to those reported in \citet{Nagai2006}.

\begin{figure*}
    \centering
    \includegraphics[width=\textwidth]{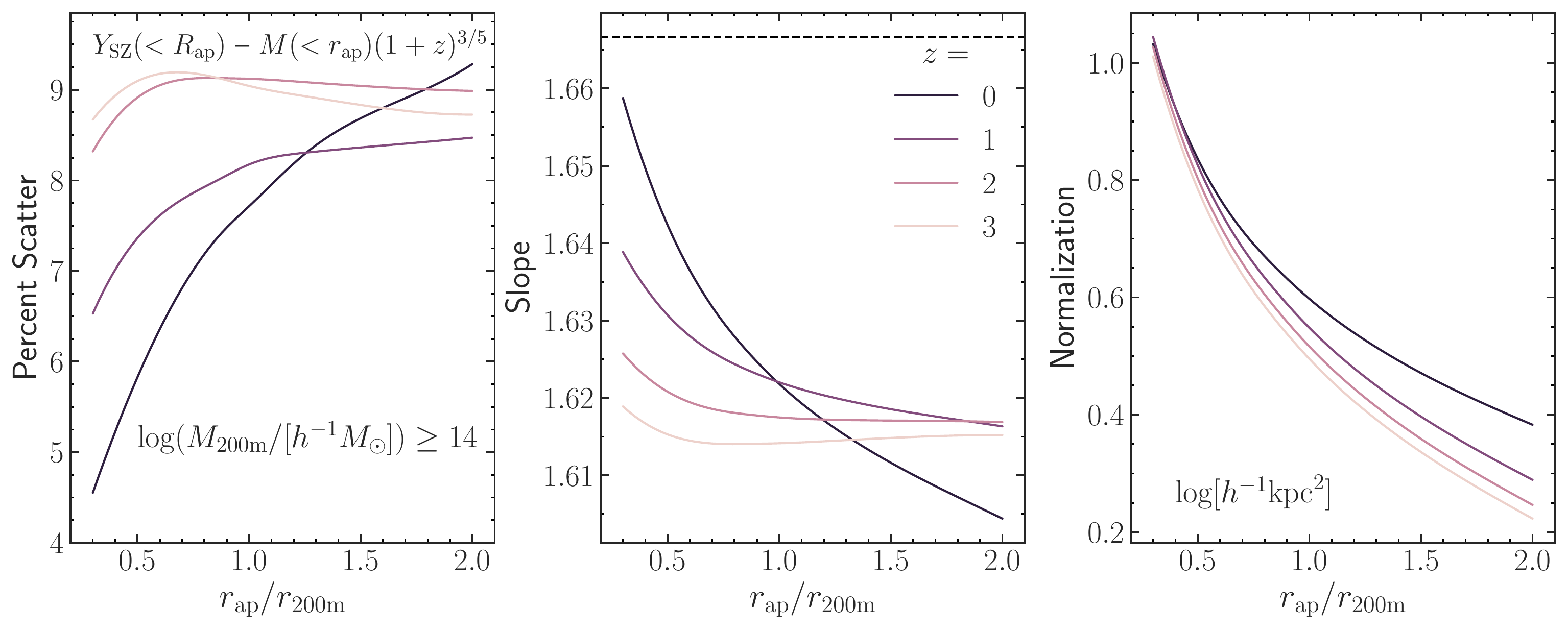}
    \caption{The best fit normalization, slope, and percent scatter of the $Y_\mathrm{SZ}(<R_\mathrm{ap}) - M(<r_\mathrm{ap})$ relation, with masses scaled by the self-similarity evolution factor, $(1+z)^{3/5}$. The fit parameters are shown as a function of $r_\mathrm{ap}$ and different curves illustrate the redshift evolution from $z=0$ to $z=3$. In these fits, {${\sim}$4,500} clusters uniformly distributed in the mass range $14 \leq \log(M_\mathrm{200m}(z) /[h^{-1}M_\odot]) \leq 15.6$ are used for each $z$. As observation redshift increases, the slope and normalization tend to decrease while the scatter increases. The interaction of the redshift-dependence of the mass-concentration relation is responsible for the apparent trend-reversals around $r_\mathrm{ap} \approx r_\mathrm{200m}$. Similar results have been seen in hydrodynamics simulations \citep[e.g.,][]{Nagai2006, Battaglia2012, LeBrun2017, Planelles2017}.}
    \label{fig:mass_obs_zeds}
\end{figure*}

\subsubsection{Redshift evolution}

Having explored the non-linear interactions between aperture radius, halo concentration, and turbulence dissipation timescales in our model, we move on to study the redshift evolution of $Y_\mathrm{SZ}-M$. In Fig.~\ref{fig:mass_obs_zeds}, we plot the best fit normalization, slope, and percent scatter for the $Y_\mathrm{SZ}-M$ scaling relation as a function of $r_\mathrm{ap}$ for different samples of clusters observed at $0 \leq z \leq 3$. When using a spherical overdensity definition relative to the mean matter density (such as $r_\mathrm{200m}$), the full self-similar scaling relation is $Y_\mathrm{SZ} \propto M^{5/3} (1+z) = [M(1+z)^{3/5}]^{5/3}$ (see Section~\ref{ssec:selfsim}). Thus, scaling the masses by $(1+z)^{3/5}$ accounts for the redshift evolution predicted by the self-similar model. Any additional redshift evolution in the normalization or slope of $Y_\mathrm{SZ}-M(1+z)^{3/5}$ indicates $z$-dependent deviations from self-similarity. The model predicts some rich trends with observation redshift. The normalization of $Y_\mathrm{SZ}-M(1+z)^{3/5}$ decreases slightly with increasing $z$, with the decrease being larger when a larger $r_\mathrm{ap}$ is used. This is simply due to the fact that at earlier times, halo MARs were generally higher (see Fig.~\ref{fig:mar_mean_std}), resulting in an overall increase in non-thermal pressure support due to turbulence, and thus suppression in $Y_\mathrm{SZ}$, with increasing $z$ (as in Fig.~\ref{fig:fnth_vs_r}). For $r_\mathrm{ap} \lesssim r_\mathrm{200m}$, the slope of the relation decreases with increasing $z$. This is due to the fact that $f_\mathrm{nth}$ in the inner regions increases more strongly with $z$ in more massive haloes (this can be seen in the left panels of Fig.~\ref{fig:fnth_vs_r}). The apparent trend-reversal at larger aperture radii is caused by the mass-concentration relation and its redshift evolution. The model also predicts that scatter in $Y_\mathrm{SZ}-M$ increases with $z$, which is directly a consequence of the increased variance in halo MARs at earlier times (see Fig.~\ref{fig:mar_mean_std}).

These redshift evolution trends are in overall agreement with predictions from NR hydrodynamical simulations, most clearly with regards to the scatter evolution. The studies by \citet{Battaglia2012}, \citet{LeBrun2017}, and \citet{Planelles2017} all find that the scatter in $Y_\mathrm{SZ}-M$ increases slightly with increasing $z$ in their NR simulations (although only for high-mass clusters in the case of \citealt{LeBrun2017}). On the other hand, only \citet{Battaglia2012} finds that the $Y_\mathrm{SZ}-M$ slope tends to decrease slightly away from self-similarity with $z$, whereas \citet{LeBrun2017} and \citet{Planelles2017} find minimal redshift evolution in the slope. The $Y_\mathrm{SZ}-M$ slope increases when going to the full-physics AGN simulations slightly in \citet{Battaglia2012} and significantly (up to ${\sim}2$) in \citet{LeBrun2017}, whereas it remains virtually unchanged in \citet{Planelles2017}, highlighting a point of tension between simulation results. These studies \citep[as well as][]{Nagai2006} have reported that the redshift evolution of the normalization shows no significant deviation from self-similarity when $r_\mathrm{ap}=r_\mathrm{500c}$, consistent with our findings for $r_\mathrm{ap} \approx 0.4 r_\mathrm{200m}$. However, the predictions of Fig.~\ref{fig:mass_obs_zeds} show that deviations from self-similarity are expected to increase in magnitude when larger aperture radii are employed. This, combined with the current tension between the results of various simulation studies (particularly with regards to the dependence of the $Y_\mathrm{SZ}-M$ slope on $z$ and AGN physics), suggests that the redshift evolution (and its dependence on $r_\mathrm{ap}$) of cluster scaling relations needs to be studied in more depth using large cluster counts. In particular, a comparison between NR and full-physics simulations will help determine whether or not the trends due to variance in MAHs predicted by our model are washed out by additional physical processes (such as AGN and supernova feedback, etc.). With upcoming surveys pushing to larger cluster counts and higher $z$, characterizing the redshift evolution of these scaling relations is of paramount importance. If our model prediction that scatter in the relations increases significantly with redshift is correct, then it will be important to continue to develop lower-scatter mass proxies with less sensitivity to redshift in order to maximally utilize upcoming high-redshift cluster data to their full potential for precision cosmology.

\begin{figure*}
    \centering
    \includegraphics[width=\textwidth]{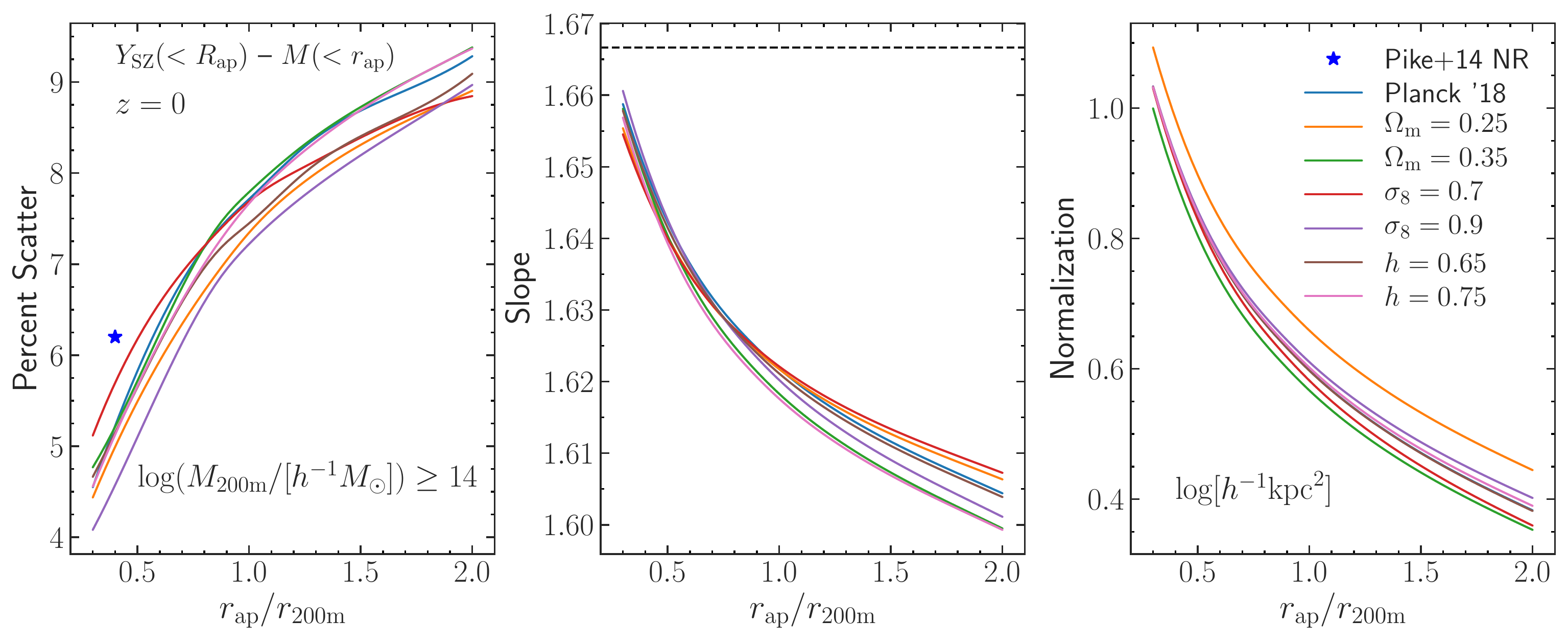}
    \caption{The best fit normalization, slope, and percent scatter of the $Y_\mathrm{SZ} (<R_\mathrm{ap})- M(<r_\mathrm{ap})$ scaling relation at $z=0$. The fit parameters are shown as a function of the aperture radius, $r_\mathrm{ap}$, in units of $r_\mathrm{200m}$. In these fits, {${\sim}$4,500} clusters in the mass range $14 \leq \log(M_\mathrm{200m}/[h^{-1}M_\odot]) \leq 15.6$ are used. Each curve represents a different cosmology, varied about the fiducial \citet{Planck18} cosmological parameters --- the variations in $Y_\mathrm{SZ}-M$ with cosmology are subtle, but the trends we find are consistent with the simulations of \citet{Singh2020}. Our predicted scatter in $Y_\mathrm{SZ}- M$ for $r_\mathrm{ap}=r_\mathrm{500c} \approx 0.4 r_\mathrm{200m}$ (at $z=0$) is only slightly below the scatter seen in the NR hydrodynamical simulations of \citet{Pike2014}. The majority of simulation studies predict scatter in the range of $10-16\%$. Hence, much of the scatter in $Y_\mathrm{SZ}- M$ is simply due to inter-cluster variation in the mass assembly histories, which drives variance in the cluster $f_\mathrm{nth}(r)$ profiles.
    }
    \label{fig:mass_obs_cosmo}
\end{figure*}

\subsubsection{Dependence on cosmology and halo mass cutoff}

In Fig.~\ref{fig:mass_obs_cosmo}, we consider the impact of single-parameter variations about the fiducial \citep{Planck18} cosmology on $Y_\mathrm{SZ}-M$ at $z=0$. Over the range of cosmologies studied, we find small but systematic trends. Recently, \citet{Singh2020} studied the effect of variations in the cosmological parameters on X-ray-based cluster scaling relations using full-physics hydrodynamics simulations based on the \textit{Magneticum} suite.\footnote{\href{http://magneticum.org/}{http://magneticum.org/}} Using an aperture of $r_\mathrm{ap} = r_\mathrm{vir}$, they find that the slope and normalization of $Y_\mathrm{SZ}-M$ systematically decrease with increasing $\Omega_\rmm$. We qualitatively reproduce these trends. While the changes to the properties of the $Y_\mathrm{SZ}-M$ relation due to large changes in the cosmological parameters (relative to the posterior distributions of the \citealt{Planck18} parameters) are small, both the present work and \citet{Singh2020} illustrate that more accurate models of the cluster scaling relations (and their dependence on cosmology) may eventually provide an additional approach to constraining the cosmological parameters given large (${\sim}10^4-10^5$) cluster samples from next-generation missions, such as \textit{eROSITA}, Simons Observatory, and CMB-S4. However, for such an approach to be feasible, future analytical gas models must account for additional significant physical processes (see Section~\ref{sec:discussion}) and the accuracy of their predictions must be validated against simulations that span a realistic range of cosmological parameters.

\new{The level of variation in the slope and scatter of the $Y_\mathrm{SZ}-M$ relation caused by changes in the cosmology are similar in magnitude to the level of variation imparted due to changing the minimum halo mass cutoff. In particular, decreasing the minimum halo mass used to compute these relations tends to decrease the overall scatter in the $Y_\mathrm{SZ}-M$ relation, since the dispersion in $\Gamma$ is lower for low-mass haloes (see Fig.~\ref{fig:mar_mean_std}). In addition, a lower mass limit tends to move the slope closer towards self-similarity due to the decrease in $f_\mathrm{nth}$ with decreasing halo mass. We find that decreasing the mass cutoff from $\log(M_\mathrm{200m}/[h^{-1}M_\odot]) \geq 14$ to $\log(M_\mathrm{200m}/[h^{-1}M_\odot]) \geq 12$ decreases the percent scatter in $Y_\mathrm{SZ}(<R_\mathrm{ap}) - M(<r_\mathrm{ap})$ by ${\sim}1\%$ (in absolute units, not relative) and increases its slope by ${\sim}0.01$ towards self-similarity. While these changes are small, this does impart a degeneracy between the mass regime used and the cosmological parameters that may become important with sufficiently large cluster samples. Importantly, we emphasize that our model does not include various physical sources of non-thermal pressure support that become increasingly important for low-mass haloes (e.g., feedback). In addition, it remains unclear how valid our choice of gas density normalization (i.e., fixed $M_\mathrm{gas}(<2r_\mathrm{200m}) / M(<2r_\mathrm{200m}) = \Omega_\rmb / \Omega_\rmm$) is for low-mass haloes. Hence, these trends with respect to the halo mass cutoff must be considered with reservation. The halo mass-dependence of $f_\mathrm{nth}$ results in a more complex relation between halo mass and observable than a simple power law can capture; future analyses should consider employing localized linear regression \citep[e.g., as used in][]{Farahi2018, Anbajagane2020} in order to quantify the mass-dependence of the scaling relation properties.}

\subsubsection{Scatter comparison with simulations and observations}
\label{sssec:scatter}

The simple model of \citetalias{Shi2014} demonstrates that a substantial fraction of the total scatter in the $Y_\mathrm{SZ}-M$ relation is likely to arise from inter-cluster variance in the non-thermal pressure, which in turn arises from variance in the halo MAHs. There have been numerous studies that address the scatter in the $Y_\mathrm{SZ}-M$ relation using simulations \citep{dSilva2004, Nagai2006, Stanek2010, Battaglia2012, Kay2012, Sembolini2013, Pike2014, Yu2015, Hahn2017, LeBrun2017, Planelles2017, Henden2019, Singh2020} as well as observations using both weak-lensing and HSE X-ray masses \citep{Bonamente2008, Hoekstra2012, Marrone2012, Planck13SZ, Czakon2015, Sereno2015, Nagarajan2019}. At $z=0$, most simulation studies find an intrinsic scatter in $Y_\mathrm{SZ}$ at fixed mass of $10-16\%$ when using $R_\mathrm{ap}=R_\mathrm{500c} \approx 0.4 R_\mathrm{200m}$. There are indications in these studies that the scatter increases slightly when going from NR runs to full-physics simulations with AGN \citep[see e.g.,][]{Battaglia2012}. On the lower end, the $Y_\mathrm{SZ}(<R_\mathrm{500c}) - M(<r_\mathrm{500c})$ relation computed using the NR simulations of \citet{Pike2014} has a scatter of just 6\%. For comparison, the intrinsic scatter in the $Y_\mathrm{SZ}-M$ relation predicted by our model, using the same aperture (see Fig.~\ref{fig:mass_obs_cosmo}) and with the same mass cutoff ($\log(M_\mathrm{200m}/[h^{-1}M_\odot]) \geq 14$), is ${\sim}5\%$. It is important to note that as SZ observation sensitivity increases, observable-mass scaling relations will be measured using larger apertures, most notably $R_\mathrm{ap}=5R_\mathrm{500c} \approx 2 R_\mathrm{200m}$. While our computation illustrates that with this larger aperture the assembly-driven scatter in $Y_\mathrm{SZ}-M$ increases to $9\%$, it is possible that contributions to the intrinsic scatter from the cluster core (largely due to feedback) will be reduced. Hence, the optimal aperture radius that minimizes the intrinsic scatter in $Y_\mathrm{SZ}-M$ is yet to be determined. \final{Additionally, our model uses a very simple prescription for the gas density profile, with its only source of halo-to-halo variance, the mass-concentration relation, introducing less than 1\% scatter into $M_\mathrm{gas}-M$. Based on the NR simulation results of \citet{Stanek2010}, we expect that by using a gas profile model that incorporates a more realistic response to halo assembly, scatter in $M_\mathrm{gas}-M$ should increase (to ${\sim}3.6\%$) and covariance between $M_\mathrm{gas}$ and $T_\mathrm{mg}$ residuals should be significant (Pearson $\rho=0.48$). Using these estimates, the scatter in our model $Y_\mathrm{SZ}-M$ should increase to $7-12\%$ (in the range of $R_\mathrm{ap}=R_\mathrm{500c}-5R_\mathrm{500c}$), which is even closer to the results of simulation studies.}

Observational studies tend to find a higher intrinsic scatter in the wider range of $14-35\%$, most of which use $R_\mathrm{ap} = R_\mathrm{500c}$, but similar results are found with $R_\mathrm{ap}= R_\mathrm{2500c}$. If a $5\%$ ($10\%$) Gaussian scatter is added to the cluster masses to mimic observational uncertainties, our predicted scatter in $Y_\mathrm{SZ}(<R_\mathrm{500c}) - M(<r_\mathrm{500c})$ increases from ${\sim}5\%$ to $10\%$ ($18\%$), which is more consistent with the observed results. The observational errors, particularly with regards to mass estimation, are still large; hence, the true intrinsic scatter in the relation is expected to be significantly lower than the values reported in the current observational literature, further motivating the development of more-accurate mass estimation techniques. However, it is also possible that additional processes not modeled in the full-physics simulations (e.g., magnetic fields and cosmic rays) are responsible for some of the additional intrinsic scatter observed.

\subsection{Mass accretion rate prediction}\label{ssec:mar}

\begin{figure}
    \centering
    \includegraphics[width=0.47\textwidth]{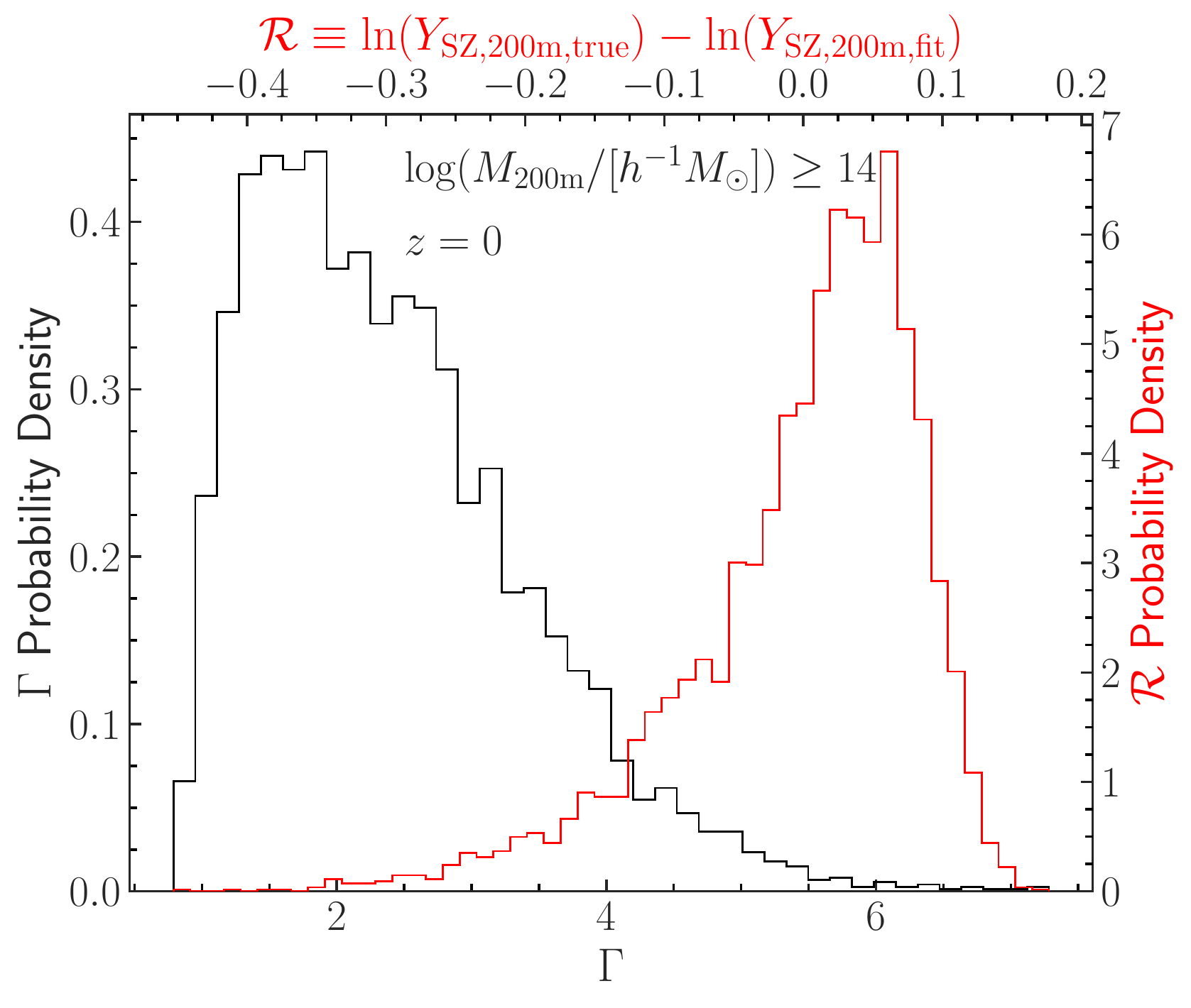}
    \caption{The distributions of halo MARs, $\Gamma$, defined by equation~\eqref{eqn:mar}, and the \new{ln-residuals}, $\mathcal{R}$, of the $Y_\mathrm{SZ}-M$ relation. The scaling relation is computed at $z=0$ for clusters in the mass range $14 \leq \log(M_\mathrm{200m}/[h^{-1}M_\odot]) \leq 15.6$. The right-skew of the MAR distribution towards a minority of disturbed clusters is responsible for the \new{left}-skew in the $\mathcal{R}$ distribution, as a high MAR increases $f_\mathrm{nth}$ and reduces the magnitude of the observables. The correspondence between the distributions suggests \new{an anti-correlation} between the two quantities (see Fig.~\ref{fig:resids_vs_mar}).}
    \label{fig:resids_and_mars_pdfs}
\end{figure}

As discussed in the Section~\ref{ssec:quant}, the model predicts a skewed distribution of the \new{ln-residuals} of the $Y_\mathrm{SZ}-M$ relation due to the skewed distribution of MARs, $\Gamma$ (see equation~[\ref{eqn:mar}]). In the \citetalias{Shi2014} model, a high recent MAR will increase $f_\mathrm{nth}(r)$, resulting in a decrease in the magnitude of the observables, $T_\mathrm{mg}$ and $Y_\mathrm{SZ}$ (at fixed halo mass). In Fig.~\ref{fig:resids_and_mars_pdfs}, we plot the distributions of $\Gamma$ and the \new{ln-residuals}, $\mathcal{R}$, computed for the $Y_\mathrm{SZ} (<R_\mathrm{200m}) - M(<r_\mathrm{200m})$ relation at $z=0$. There is a strong \new{left}-skew in the $\mathcal{R}$ distribution towards over-predictions, and this skewness is mirrored in the MAR distribution towards a small fraction of haloes with high $\Gamma$ (i.e., disturbed clusters). The skewness in $\mathcal{R}$ is present regardless of mass cutoff or aperture employed. 

\begin{figure*}
    \centering
    \includegraphics[width=\textwidth]{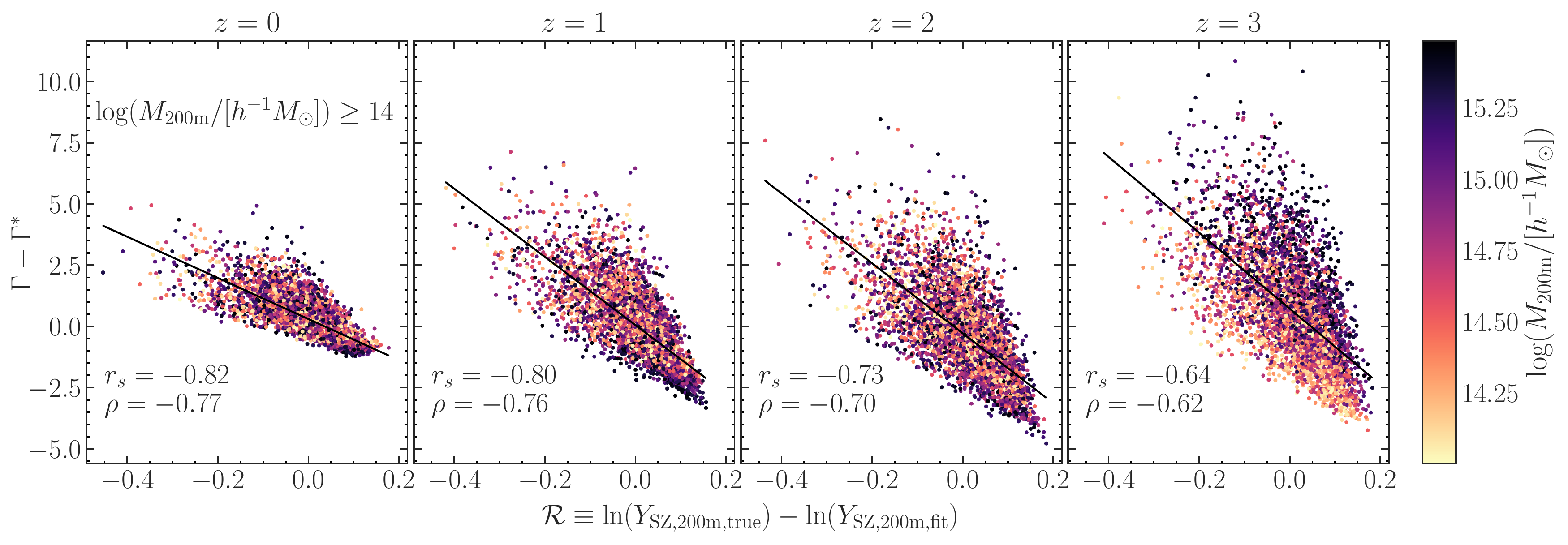}
    \caption{The relationship between the MARs and the $Y_\mathrm{SZ}-M$ \new{ln-residuals} for clusters at $0\leq z \leq 3$ and in the mass range $\log(M_\mathrm{200m}/[h^{-1}M_\odot]) \geq 14$. \new{The MARs are plot relative to the median MAR of haloes of a given $z$ and $\nu_\mathrm{200m}$, denoted $\Gamma^*$, using the fitting function of \citet{Diemer2017b}.} At higher $\Gamma$, a halo tends to have more non-thermal pressure, which reduces the magnitude of $Y_\mathrm{SZ}$, ultimately \new{decreasing} $\mathcal{R}$. \new{The trend is strongest at $z=0$, with a Pearson $\rho=-0.77$ and Spearman $r_s=-0.82$, weakening slightly as $z$ increases. The slope in the relation tends to steepen with $z$.} If present in full-physics simulations, this $\Gamma-\mathcal{R}$ relation may provide a link between the splashback radius and the \textit{observable} residual, $\mathcal{R}$.}
    \label{fig:resids_vs_mar}
\end{figure*}

The correspondence between the two distributions suggests that the $Y_\mathrm{SZ}-M$ residual, which is itself an observable quantity, is likely to \new{anti-}correlate with the underlying halo MAR. The ability to estimate $\Gamma$ from an observable would be powerful since, as discussed in \citet{Diemer2017b}, $\Gamma$ is closely connected to the splashback radius and mass, $r_\mathrm{sp}/r_\mathrm{200m}$ and $M_\mathrm{sp}/M_\mathrm{200m}$. The splashback radius has been suggested as a better, physically-motivated definition for the halo boundary \citep{Adhikari2014, Diemer2014, More2015, Mansfield2017, Xhakaj2019}, but it has proven difficult thus far to observe $r_\mathrm{sp}$ for individual clusters.

In Fig.~\ref{fig:resids_vs_mar}, we plot the cluster MARs against their $Y_\mathrm{SZ}-M$ \new{ln-residuals} for several different redshifts. \citet{Diemer2017b} present a fitting function for the median MAR seen in cosmological simulations as a function of $z$ and $\nu_\mathrm{200m}$, which we denote $\Gamma^*$ and \new{use to standardize our $\Gamma$ values (i.e., $\Gamma-\Gamma^*$).} There is a strong trend between the scaling relation \new{ln-residual} and the median-\new{standardized MAR, with a Pearson $\rho=-0.77$ and Spearman $r_s=-0.82$ at $z=0$.} The slope of the relation tends to \new{increase in magnitude} slightly with $z$. Importantly, for $\mathcal{R}=0$, the trend predicts that $\Gamma - \Gamma^* \approx 1$; in other words, if the cluster falls on the best-fit line for the $Y_\mathrm{SZ}-M$ relation, its MAR tends to be around the median for a halo of its mass at $z$. It is unclear whether or not such a strong trend between the residuals and halo MAR exists in real clusters, since previous simulation studies \citep[e.g.,][]{Battaglia2012} have found that the MAR-driven skewness in the $Y_\mathrm{SZ}-M$ residuals decreases and the distribution approaches normality when additional physics beyond NR hydrodynamics is modeled in the simulations. The relationship between the observable $Y_\mathrm{SZ}-M$ \new{ln-residuals} and $\Gamma$ should be explored in future full-physics simulation studies in order to quantitatively measure the strength of the $\Gamma-\mathcal{R}$ relation and forecast its predictive power for determining other secondary cluster properties that have recently been tied to the MAR, such as assembly bias \citep{Sunayama2019} and the asphericity of the ICM \citep{Chen2019}.

\section{Discussion}\label{sec:discussion}

Our approach assumes that non-thermal pressure is dominated by turbulence generated during mass assembly. The \citetalias{Shi2014} model of the non-thermal pressure profile does not yet take into account various secondary effects due to baryonic physics, many of which will likely increase the intrinsic scatter in the scaling relations from what is presented here, especially for low-mass haloes and when small apertures are used. Radiative cooling and star formation results in the condensation of gas into the center of the cluster, reducing the baryon budget. Both \citet{Shaw2010} and \citet{Flender2017} modeled this by assuming that the gas adiabatically contracts or expands due to the change in total gas mass. Feedback due to AGN and supernovae provide additional sources of turbulence, especially in the inner regions of the cluster \citep[e.g.,][]{Vazza2013, Zhuravleva2014, Chadayammuri2020}. These feedback effects become more significant as the halo mass decreases; hence, they must be accurately modeled in order to successfully quantify the thermodynamic properties of low-mass haloes through stacked measurements from upcoming microwave and X-ray surveys. 

\new{Observations of non-thermal X-ray emission and radio haloes \citep[e.g.,][]{Million2009, Kale2013} imply the presence of additional non-thermal pressure due to cosmic rays and magnetic fields in the ICM. Substantial turbulent energy can also be injected into the cluster outskirts by the magneto-thermal instability \citep{Parrish2008, Parrish2012}.} Strong constraints have been placed on the magnetic field strength in the ICM, limiting the magnetic field-associated pressure to be much smaller than the thermal pressure \citep[$\approx 5\%$,][]{Dolag2000, Iapichino2012}. Observations of $\gamma$-ray emission in nearby clusters provide constraints on the pressure due to cosmic ray protons generated from shocks in the ICM to be less than 2$\%$ of the thermal pressure \citep{Ackermann2014, Shirasaki2019}. Some simulations, however, suggest that cosmic rays could provide almost $50\%$ of the total pressure in the cluster cores \citep{Sijacki2008}. Thus, although the overall additional non-thermal pressure due to magnetic fields and cosmic rays is likely small, better constraints are still warranted in order to determine the importance of incorporating their effects into future models of the total non-thermal pressure support. 

Throughout our work, we assume spherically symmetric pressure and gas profiles for the clusters. The observational analysis of \citet{Arnaud2010} has shown that deviations from spherical symmetry and variations in cluster shapes can lead to scatter in the spherically-averaged pressure profiles. The recent hydrodynamical simulation study of \citet{Chen2019} has also reported that deviations from spherical symmetry increase the scatter in X-ray-based observable-mass scaling relations, additionally illustrating that the ellipticity of the ICM may be seeded by the MAH. Hence, the impact of mass assembly on the scatter in both SZ and X-ray observable-mass relations studied using our approach should still be regarded as a lower bound. The strength of future theoretical models will be greatly increased by incorporating the effect of mass accretion on triaxiality and cluster shape.

We have also neglected the impact of line-of-sight projection effects on the cluster observables. In particular, we study some spherically-integrated observables (i.e., $M_\mathrm{gas}$ and $T_\mathrm{mg}$) and cut off the $y_\mathrm{SZ}(R)$ line-of-sight integration at $2r_\mathrm{200m}$. However, simulated light cone studies have demonstrated that a non-negligible fraction of the SZ signal arises from from the warm diffuse gas residing outside of groups and clusters \citep{Hallman2007}. Furthermore, \citet{Shirasaki2016} found that the projection of correlated structures along the line-of-sight introduces additional scatter into the scaling relation between the tSZ effect signal and the weak lensing mass. Thus, future gas models that aim to be combined with $N$-body simulations for efficient production of mock light cones must take into account the impact of the warm-hot intergalactic medium and other correlated structures along the line-of-sight.

\section{Conclusion and Future Outlook}\label{sec:conclusion}

We quantified the effects of mass assembly-driven turbulence on the $Y_\mathrm{SZ}-M$ scaling relation. This was accomplished by combining a simple model of the total pressure and gas density profiles, a model of the evolution of non-thermal pressure, and Monte Carlo-generated halo mass accretion histories. We summarize our most salient findings below:
\begin{itemize}
    \item The average non-thermal pressure fraction profiles, $f_\mathrm{nth}(r)$, tend to increase as halo mass or observation redshift increases. This is simply due to the fact that (i) higher mass haloes assemble later and (ii) a higher redshift of observation requires more rapid mass accretion at fixed halo mass.
    \item When radii are normalized by $r_\mathrm{200m}$, the model predicts $f_\mathrm{nth}(r/r_\mathrm{200m})$ profiles that exhibit near-universality in redshift \textit{at fixed peak height}, $\nu_\mathrm{200m}$. This finding is consistent with the simulation study of \citet{Nelson2014}. We provide a fitting function for $f_\mathrm{nth}(r/r_\mathrm{200m} \vert \nu_\mathrm{200m})$ described by equation~\eqref{eqn:fnth_fit} and Table~\ref{tab:fit}.
    \item As a consequence of $f_\mathrm{nth}(r)$ increasing with halo mass and redshift, the model predicts that the magnitude of the average HSE mass bias (i.e., the deviation of the HSE-inferred mass from the true mass) also experiences these same trends.
    \item The scatter in the $Y_\mathrm{SZ}-M$ relation due solely to inter-cluster variance in the halo MAHs ranges from $5-9\%$, increasing with aperture radius and $z$. \final{This should be regarded as a lower bound, as the scatter will likely increase by a few percent once a more realistic model of the gas density profile is incorporated.} For reference, most NR hydrodynamical simulations predict $10-15\%$ scatter. Thus, our model predicts that assembly-driven turbulence is responsible for a substantial fraction of the total scatter in $Y_\mathrm{SZ}-M$.
    \item The slope of $Y_\mathrm{SZ}-M$ tends to decrease slightly away from the self-similarity slope of $5/3$ as aperture, redshift, or halo mass limit increases. This dependence on aperture was also reported in \citet{Nagai2006}.
    \item There are small trends in the slope, normalization, and scatter of $Y_\mathrm{SZ}-M$ with cosmology. The trends seen in Fig.~\ref{fig:mass_obs_cosmo} are consistent with those seen in the X-ray observable-mass relations of the simulated clusters in \citet{Singh2020}. The perturbations in $Y_\mathrm{SZ}-M$ due to cosmology are similar in magnitude to those seen due to variations in the lower mass cutoff used for computing the relation. This indicates that careful control of sample selection will be essential for any attempt to use cluster scaling relations to constrain cosmological parameters.
    \item The model predicts a skewed distribution of \new{ln-residuals}, $\mathcal{R}$, for $Y_\mathrm{SZ}-M$ due to the skewed distribution of $\Gamma$, in agreement with the NR hydrodynamical simulations of \citet{Battaglia2012}. We find that $\Gamma$ \new{anti-}correlates tightly with $\mathcal{R}$ (Spearman $r_s=-0.82$ at $z=0$), potentially introducing a new observational approach to estimating the mass accretion rate via $Y_\mathrm{SZ}-M$. 
\end{itemize}

The non-thermal pressure support present in galaxy clusters must be taken into account in order to make accurate HSE mass estimates and utilize the full statistical power that will be available in next-generation X-ray and SZ surveys for cluster count-based cosmological analyses. By studying the non-thermal pressure fraction profile, which is an important component of analytical models of the ICM \citep{Shaw2010,Flender2017}, we highlighted the dependence of accretion-driven turbulence on halo mass and redshift. As survey sensitivity continues to grow, the need to model and correct for the HSE mass bias over a wide range of halo masses and redshifts (especially smaller group mass haloes and high-redshift systems) is becoming increasingly important. This work represents a step towards developing a more accurate analytical model of the hot gas in groups and clusters, which will help (i) disentangle the effects of AGN/supernovae feedback from the non-thermal pressure driven by the structure formation process and (ii) model the cosmological dependence of the ICM. 

\final{The current model of the gas density, developed in \citetalias{Komatsu.Seljak2001}, is very simple and does not include any baryonic physics. Hence,} a promising next step in model development should involve incorporating the effects of galaxy formation physics into the dark matter and gas density models \citep[e.g.,][]{Schneider2020}, which would enable the modeling of both galaxy formation and structure formation physics in a unified analytical framework. \new{In addition, future cosmological simulations should focus on illuminating the importance of additional physical effects. Idealized simulations may over-predict the non-thermal pressure attributed to magnetic fields and thermal conduction \citep{Parrish2012} since the turbulence and shocks generated by the structure formation process interact non-linearly with magnetic fields, which can lead to turbulence that changes non-monotonically with halo mass \citep{McCourt2013}. Modeling the turbulence pressure caused by additional sources as well as capturing baryonic effects on the dark matter (i.e., halo response modeling) and ultimately calibrating such models based on simulations} will be crucial for combining the ICM model with models of the galaxy-halo connection and $N$-body simulations to generate a physically motivated and computationally efficient framework for interpreting forthcoming multi-wavelength cosmological datasets. Such an approach will eventually enable the use of correlation statistics from multi-wavelength cosmological surveys to constrain cosmology and astrophysics \citep{Shirasaki2020}.

Currently, the best observations of bulk and turbulent motions in the ICM are of the Perseus cluster core, where the \textit{Hitomi} X-ray observatory has reported high-resolution measurements of emission line Doppler shifting and broadening \citep[][for a recent review]{Hitomi2018, Simionescu2019}. In the near future, \textit{XRISM}/Resolve \citep{XRISM} and \textit{Athena}/X-IFU \citep{ATHENA} will measure the turbulence in the ICM for many nearby clusters and within the cores of more distant clusters, providing an opportunity to check the $f_\mathrm{nth}(r)$ model, correct for the HSE mass bias, and properly calibrate the mass scale \citep{Ota2018}. Furthermore, the \textit{Lynx} X-ray Surveyor \citep{Gaskin2019} and \textit{Cosmic Web Explorer} \citep{Simionescu2019b} have been proposed as future-generation X-ray telescopes that would enable exquisite measurements of turbulence out to the halo outskirts of an unprecedentedly large sample down to the galaxy mass scale.

In the future, millimetre-wave observations may provide a promising and complementary lens into the thermodynamics of and gas motions in the ICM via the thermal and kinematic SZ effects \citep[see e.g.,][]{Mroczkowski2019}. Upcoming and proposed microwave instruments, such as the TolTEC camera,\footnote{\href{http://toltec.astro.umass.edu/}{http://toltec.astro.umass.edu/}} CCAT-prime,\footnote{\href{http://www.ccatobservatory.org/}{http://www.ccatobservatory.org/}} CMB-HD \citep{Sehgal2019}, and Voyage2050 \citep{Basu2019b}, will enable high-resolution SZ spectral imaging of clusters. This additional spectral information encodes a measurement of the kinematic SZ effect, which can be used to separate the cluster peculiar velocity and internal velocity dispersion \citep{Inogamov2003,Nagai2003, Sayers2019}, thus providing a direct measurement of the non-thermal pressure support. Furthermore, since the strength of the SZ signal is independent of redshift, this approach can be used to observe the redshift-dependence of $f_\mathrm{nth}(r)$. Lastly, these observations will facilitate relativistic SZ corrections, which can be leveraged to study temperature structures in the ICM and mitigate the biases in the derived SZ and X-ray temperatures \citep[see e.g.,][]{Chluba2012, Chluba2013, Lee2020}.

Finally, previous attempts at measuring the mass accretion rate of clusters via its relationship to the splashback radius have suffered from systematic uncertainties such as selection and projection effects \citep{Baxter2017, Busch2017, Zu2017}. The strong correlation between $\Gamma$ and the \new{ln-residuals}, $\mathcal{R}$, of the $Y_\mathrm{SZ}-M$ relation highlighted in this study may provide an alternative means to measure the MAR, provided that the relationship is not washed out by other sources of non-thermal pressure or by observational errors. In addition to this $\Gamma-\mathcal{R}$ relation, machine learning algorithms may provide an alternative approach that enables more accurate determinations of both $\Gamma$ and the cluster mass, employing input features such as images of the ICM and summary statistics that quantify the cluster shape \citep[e.g.,][]{Green2019, Ntampaka2019}. 

\section*{Acknowledgements}

\new{The authors thank the referee, August Evrard, for his useful report that helped improve this manuscript. We also thank Xun Shi for useful comments on the manuscript.} SBG is supported by the US National Science Foundation Graduate Research Fellowship under Grant No. DGE-1752134. DN acknowledges support by National Science Foundation grant AST-1412768 and the hospitality at the Aspen Center for Physics, which is supported by National Science Foundation grant PHY-1607611. FCvdB is supported by the National Aeronautics and Space Administration through Grant Nos. 17-ATP17-0028 and 19-ATP19-0059 issued as part of the Astrophysics Theory Program and received additional support from the Klaus Tschira foundation. This work was supported in part by the facilities and staff of the Yale Center for Research Computing.

\section*{Data Availability}

The code used to generate the average halo mass accretion histories is available online.\footnote{\href{http://www.astro.yale.edu/vdbosch/PWGH_files/getPWGH.tar}{http://www.astro.yale.edu/vdbosch/PWGH\_files/getPWGH.tar}} The code used to generate Monte Carlo halo MAHs, as well as all other analysis and plotting routines, is available upon request from the corresponding author.

\bibliographystyle{mnras}
\bibliography{references}

\bsp	
\label{lastpage}
\end{document}